\documentclass[%
 reprint,
 superscriptaddress,
 amsmath,amssymb,
 aps,
 prb,
]{revtex4-2}

\usepackage{subfigure}
\usepackage{graphicx}% Include figure files
\usepackage{dcolumn}% Align table columns on decimal point
\usepackage{bm}% bold math
\usepackage{hyperref}% add hypertext capabilities
\hypersetup{
    colorlinks=true,
    linkcolor=blue,
    filecolor=magenta,      
    urlcolor=cyan,
}
\usepackage{balance}
\usepackage{float}
\usepackage{soul}
\usepackage{braket}

\usepackage[normalem]{ulem}
\usepackage[dvipsnames]{xcolor}

\newcommand*{\balancecolsandclearpage}{%
  \clearpage
  \twocolumngrid
}

\begin{document}
% \linenumbers

% \preprint{APS/123-QED}

\title{Phase Diagram Detection via Gaussian Fitting of Number Probability Distribution}
%Full-counting statistics: a simple phase diagram detector

\author{Daniele Contessi}
\affiliation{Dipartimento  di  Fisica,  Universit\`a  di  Trento \&  INO-CNR BEC Center,  38123  Povo,  Italy;}
\affiliation{Forschungszentrum J\"ulich GmbH, Institute of Quantum Control,\\
Peter Gr\"unberg Institut (PGI-8), 52425 J\"ulich, Germany; }
\affiliation{Institute for Theoretical Physics, University of Cologne, D-50937 K\"oln, Germany}

\author{Alessio Recati}
\affiliation{Dipartimento  di  Fisica,  Universit\`a  di  Trento \&  INO-CNR BEC Center,  38123  Povo,  Italy;}

\author{Matteo Rizzi}
\affiliation{Forschungszentrum J\"ulich GmbH, Institute of Quantum Control,\\
Peter Gr\"unberg Institut (PGI-8), 52425 J\"ulich, Germany; }
\affiliation{Institute for Theoretical Physics, University of Cologne, D-50937 K\"oln, Germany}

\date{February 2023}

%%%%%%%%%%%%%%%%%%%%%%%
%%% ABSTRACT
%%%%%%%%%%%%%%%%%%%%%%%

\begin{abstract}
We investigate the number probability density function that characterizes sub-portions of a quantum many-body system with globally conserved number of particles. 
We put forward a linear fitting protocol capable of mapping out the ground-state phase diagram of the rich one-dimensional extended Bose-Hubbard model:
The results are quantitatively comparable with more sophisticated traditional and machine learning techniques.
We argue that the studied quantity should be considered among the most informative bipartite properties, being moreover readily accessible in atomic gases experiments.    
\end{abstract}

\maketitle

%%%%%%%%%%%%%%%%%%%%%%%
%%% INTRODUCTION
%%%%%%%%%%%%%%%%%%%%%%%

% \begin{figure}[hbt]

%         \includegraphics[width = 0.45\textwidth]{./PBClattice.pdf}

%     \caption{\label{fig:sketch}(color online).  
%     }
% \end{figure}
%

%\Teo{per ora scrivo senza referenze, poi aggiustiamo}

\section{Introduction:}

In recent years, quantum information and condensed matter theory have considerably cross-fertilized, and a wealth of properties (mainly) of bipartite systems have emerged as powerful indicators of the many-body wave-function properties.
With their help, quantum phases of matter and transitions between them have been characterized, and put in relation with the underlying quantum field theories.
When dealing with the reduced density matrix of a sub-system, concepts like the  entropy of the associated probability distribution, as well as its spectrum (technically the Schmidt singular values), are indeed unveiling the entanglement properties of the state of the system. 
A first prominent example is the log-scaling of the von Neumann (and, more generally, of any R\`enyi) entanglement entropy with the bipartition size for  one-dimensional critical systems with local Hamiltonians~\cite{Vidal2003,CalabreseCardy,LeHurEE}: Fitting the coefficient in front of such law is arguably one of the best ways to estimate the so-called central charge $c$ of the associated conformal field theory (CFT). 
As a counterpart, the entanglement spectrum (ES) itself has been proven to exhibit peculiar properties both in gapless and gapped phases~\cite{CalabreseLefevre,lauchli_gapless,lauchli_gapped}, and its degeneracy pattern is often used to tell different kinds of topological phases apart~\cite{HaldaneQuantumHall,Pollmann_2010}. 
Its structures embed information about non-local quantum correlations, as formalized within  the bulk-boundary correspondence framework~\cite{BulkEdgeRegnault2011,WenBookQFT,WenBookQIQMB}. %\Dani{Check Wen book}
Both ad-hoc defined order parameters~\cite{Santos2011,SanperaSchmidtGap} and machine learning driven approaches~\cite{EvertNature2017,Maciej2021,our_paper_GAN} have been employed for the detection of phase transitions.
Last, but certainly not least, the evolution of entanglement properties under the system dynamics has recently unveiled the existence of new kinds of non-equilibrium phase transitions for quantum many-body systems under random projective measurements or unitary gates~\cite{RandomQuantumCircuits2017,EntanglementPhaseTransition2019Nahum,EntanglementPhaseTransition2019Fisher}.

As a consequence of the above, considerable efforts have been spent towards making such entanglement features measurable in the laboratory~\cite{MeasureEEZoller2012,MeasureEEGreinerNature2015,MeasureEEGreinerScience2016,MeasureRenyiZollerScience2019,MarcelloBisognanoWichmann2018}.
%
%\Teo{\emph{(Marcello \& co. con Bisognano-Wichmann, \ldots)}}\Dani{Dici questo o roba più vecchia? \cite{MarcelloBisognanoWichmann2018}}
%\Teo{\emph{(mi sembra giusto, se non hanno PRL o altro)}}
%
Among the many, the so-called number entanglement entropy is gaining a prominent role:
Its operational definition refers to the probability density function (PDF) of a U(1) globally conserved charge in an extensive sub-portion of the system,
also for mixed states~\cite{NumberEntropy2022}. Evidences of its distinctive dynamical behaviour as a hallmark for many-body localization have been recently obtained numerically and experimentally~\cite{ GreinerMBL2019,KieferPRL2020,KieferAnnalsOfPhysics2021}.
%\Teo{\emph{(sarebbe bene magari una frase ``di concetto'')}}

A specific property of the PDF has been explored in the past, namely its second momentum or, in other words, the amount of charge fluctuations $\mathcal{F}$ across sub-systems. 
As very extensively explained in~\cite{LeHurLong}, a lot of properties of the fluctuations are shared with the entanglement entropy: for a gapped phase, they exhibit a strict area-law behaviour, $\mathcal{F}\propto L^{d-1}$, with $L$ the linear size of the partition and $d$ the dimension of the system, whereas for a gapless phase there appears a logarithmic correction: $\mathcal{F}\propto L^{d-1}\ln L$~\cite{LeHurQCP}.
In particular for a Luttinger liquid, the scaling coefficient is related to the $K$-parameter, thus yielding yet another piece of information about the underlying field theory.

%\Prev{In this Letter, we propose to look at the full probability density function as a valuable tool to map out the phase diagram of quantum many-body systems, in a similar fashion to what has been successfully performed with the entanglement spectrum.
%As we will detail below, indeed, the two quantities are intimately related to each other, with the PDF being a very informative reduction of the ES readily measurable, e.g., in quantum gas microscope experiments.
%We argue that a simple and very general procedure consisting only of some linear fits of the PDF is sufficient to read out a wealth of phases with minimal prior knowledge about their nature. 
%We illustrate that analysing the one-dimensional extended Bose-Hubbard (EBH) model at zero temperature.
%Indeed, some functional forms of the PDF for limiting scenarios have been put forward in previous literature~\cite{LevitovKlichPRL2012,LeHurEE,EranPRL2018,CalabreseBonsignoriFreeFermi2019,CalabreseCapizzi2020,CalabreseXXZFullCountingStatistics2020}: 
%Here we show that they can be connected to each other.}
In this Letter, we propose a new approach to map out the phase diagram of quantum many-body systems by considering the full PDF. 
We are able to detect all the phase transitions of the one-dimensional extended Bose-Hubbard (EBH) model at zero temperature by performing a simple and yet very general procedure consisting of some educated fits of the PDF only, therefore without resorting to any phase-specific quantity like order parameter or correlation. 
We show how the PDF, being intimately related with the ES, preserves its intrinsic wealth of information about the nature of the phases. 
This can be exploited for an agnostic and automatic detection, as done with the machine learning solutions to the problem involving the ES.
The huge advantage of the full PDF is its availability in experiments, e.g. with quantum gas microscopes, and not only in numerical simulations like the ones we perform here via matrix-product-states (MPS) simulations with embedded quantum numbers.
We also show that a finite-size scaling analysis of the PDF leads to a pretty precise determination of the phase boundaries, 
and that some previously found functional forms of the PDF for limiting scenarios~\cite{LevitovKlichPRL2012,LeHurEE,EranPRL2018,CalabreseBonsignoriFreeFermi2019,CalabreseCapizzi2020,CalabreseXXZFullCountingStatistics2020} can be connected to each other.

%We conclude with an outlook on possible generalizations to other models and higher-dimensional systems.
%\Teo{\emph{(da aggiustare a seconda di cosa poi davvero scriviamo)}}

%%%%%%%%%%%%%%%%%%%%%%%%%%%%%%%%%%%%%%%%%%%
%%% MODEL
%%%%%%%%%%%%%%%%%%%%%%%%%%%%%%%%%%%%%%%%%%%

\section{Model \& Method:}
The Hamiltonian of the one-dimensional EBH model can be written as:
\begin{equation}
%    \begin{split}
%        H = &-t \sum_{j=1}^{L}(b^\dagger_{j+1}b_j + \mathrm{h.c.}) \\&+ \frac{U}{2}\sum_{j=1}^{L} n_j(n_j-1) + V\sum_{j=1}^{L} n_j n_{j+1},
%    \end{split}
        H = \sum_{j=1}^{L} \left( - t (b^\dagger_{j+1}b_j + \mathrm{h.c.}) + \frac{U}{2} n_j(n_j-1) + V n_j n_{j+1} \right),
    \label{eq:Hamiltonian}
\end{equation}
where $b^{(\dagger)}_j$ is the annihilation (creation) bosonic operator and $n_j = b^{\dagger}_j b_j$ the number operator on site $j$ along a one-dimensional chain with periodic boundary conditions (PBC). The hopping coefficient is denoted by $t$, while $U$ and $V$ are the on-site and nearest-neighbour interaction strengths. 
The zero-temperature phase diagram of the unitary integer filled lattice, $\nu = N/L = 1$, presents a number of prototypical quantum phase transitions~\cite{fazio2012, Holger2014HI}:
(i) a Berezinskii-Kosterlitz-Thouless (BKT) transition between a gapless superfluid (SF) and a gapped Mott insulator (MI) phases, 
(ii) a $c=1$ transition between the MI and a topological Haldane insulator (HI),
and (iii) an Ising $c=1/2$ transition between the HI and a charge density-wave (CDW) state.
%The (i) transition belongs to the Berezinskii-Kosterlitz-Thouless (BKT) universality class, the (ii) transition is in the Luttinger liquid universality class, while the last transition (iii) belongs to the Ising universality class.  
Moreover, it has been recently pointed out that a phase-separated regime between SF and supersolid (SF + SS)~\cite{Maciej2021,macejiSS} is present in the bottom right corner of the phase diagram.
We set the energy scale by taking $t=1$, and focus on the region $U\in[0,6]$ and $V\in[0,5]$ in order to have a direct comparison with the works in~\cite{fazio2012,Maciej2021}.

We determine the ground state $\vert \psi_g \rangle$ of the EBH model for various parameters by means of a Matrix Product States (MPS) ansatz with U(1) symmetric tensors. 
Indeed the EBH model is non-integrable and our numerical treatment is an almost unbiased approach to it.
We set the maximum occupation to $n_{max}=4$ bosons per site.
We deal with PBC by employing a loop-free geometry of the tensor network and shifting the topology of the lattice into the Matrix Product Operator (MPO) representation of the Hamiltonian (see Supplementary Material of~\cite{ABpaper} for a detailed description).
In this way, we can reliably compute relevant quantities for systems up to $L=256$ sites, with a discarded probability not exceeding $10^{-9}$. 
%a bond dimension up to $\chi_{max}=400$ and 

%The Tensor Network approach we use is particularly suitable to calculate the PDF of a portion of the system. 
We consider the system as divided in two equal portions $A$ and $\bar{A}$: 
While the total number of particles is conserved, the number in the single region can fluctuate. 
The measurement of a deviation $\delta n = m$ from the average density happens with a probability 
\begin{equation}
    p(\delta n = m) = \mathrm{Tr}(\rho_A\Pi_{N/2 +m}) = \sum_{\alpha} \lambda_{\alpha}^{(m)},
    \label{eq:pm}
\end{equation}
where $\rho_A := \mathrm{Tr}_{\bar{A}} \vert \psi_g \rangle \langle \psi_g \vert$ is the reduced density matrix of sub-system A, and $\Pi_{N/2 +m}$ is the projector on the sector with $N/2 +m$ occupancy.
In the last equality we introduced the eigenvalues $\lambda_\alpha^{(m)}$ of $\rho_A$ in the $N_A = N/2+m$ particles sector: 
They are also related to so-called ES eigenvalues $\xi_\alpha^{(m)}$, via the relation $\lambda_\alpha^{(m)} = e^{-\xi_\alpha^{(m)}}$~\cite{HaldaneQuantumHall,Pollmann_2010}. 
Being the $\lambda_\alpha^{(m)}$s the natural metric on which truncations of the Tensor Network representation are performed, the PDF is automatically at disposal without extra computational costs.
We notice here that, in an experimental setup with access to site-resolved populations, the PDF is obtained by bin-counting the occupation numbers in half of the system~\cite{GreinerMBL2019}.

%%%%%%%%%%%%%%%%%%%%%%%
%%% GAUSSIAN PDF
%%%%%%%%%%%%%%%%%%%%%%%

%%%%%%%%%%%%%%%%%%%%%%%%%%%%%%%%%%%%%%%%%%%%%%%%%%%%%%%
\begin{figure}[t]
\centering
    %\subfigure{
        \includegraphics[width = 0.5\textwidth]{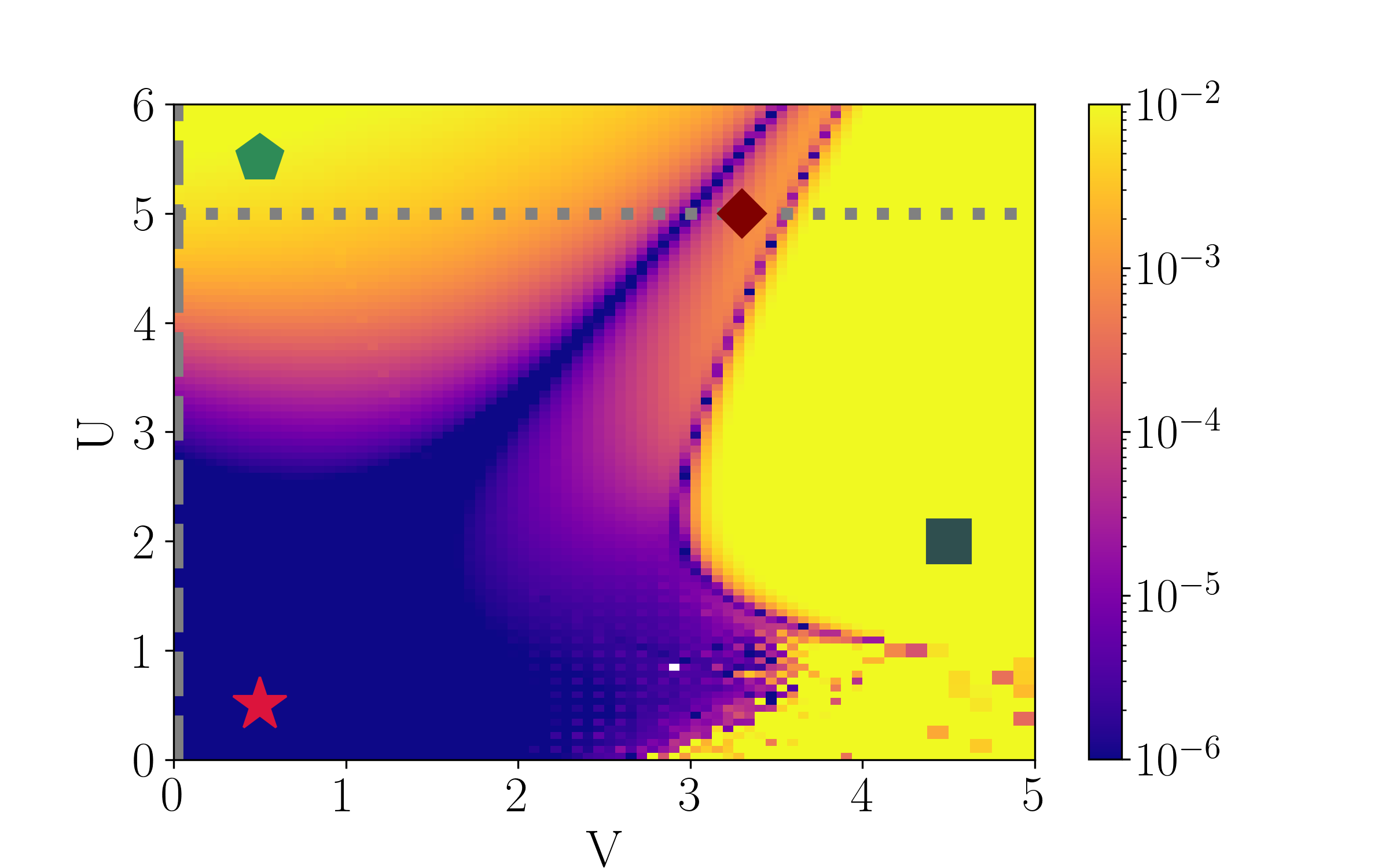}%}\hspace{-0.5in}
    % \subfigure{
    %     \includegraphics[width = 0.5\textwidth]{./gapped_diagram_clean.png}
    % }
    \caption{\label{fig:phase_diagram}(color online). Phase diagram of the EBH model traced with the residuals of the PDF's quadratic fit $p(m)\propto e^{- \beta m^2}$ for a $L=64$ system. Details about the phase transitions along the dashed and dotted cuts are shown in Fig.~\ref{fig:cuts}. The PDFs correspondent to the configurations marked with the colored shapes are displayed below in Fig.~\ref{fig:PDFdetails}.}
\end{figure}
%%%%%%%%%%%%%%%%%%%%%%%%%%%%%%%%%%%%%%%%%%%%%%%%%%%%%%%

%%%%%%%%%%%%%%%%%%%%%%%%%%%%%%%%%%%%%%%%%%%%%%%%%%%%%%%%
\begin{figure*}[hbt]
\centering
    \subfigure{
        \includegraphics[width = 0.5\textwidth]{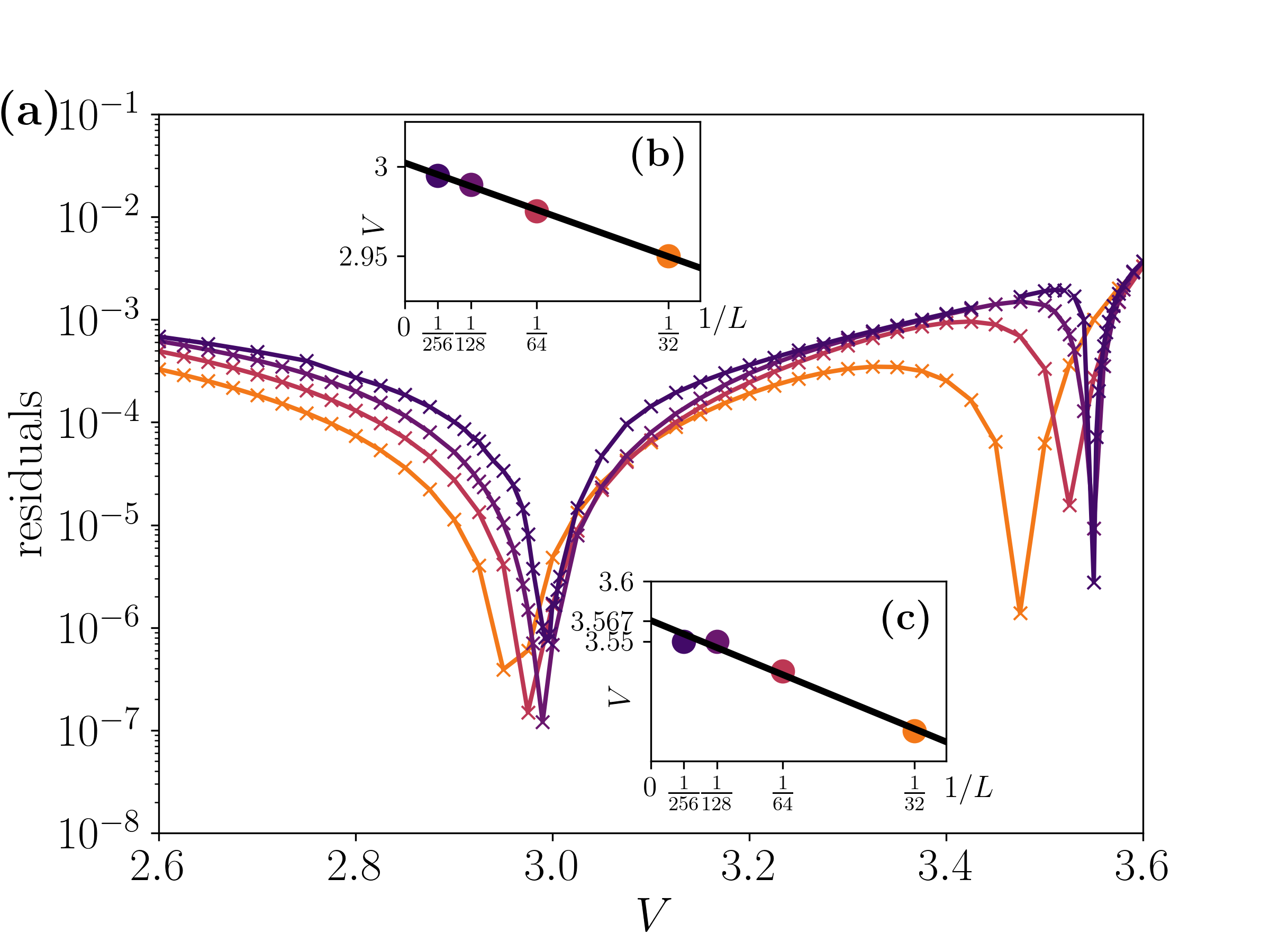}%
        \hspace{-0.25in}
        }
    \subfigure{
        \includegraphics[width = 0.5\textwidth]{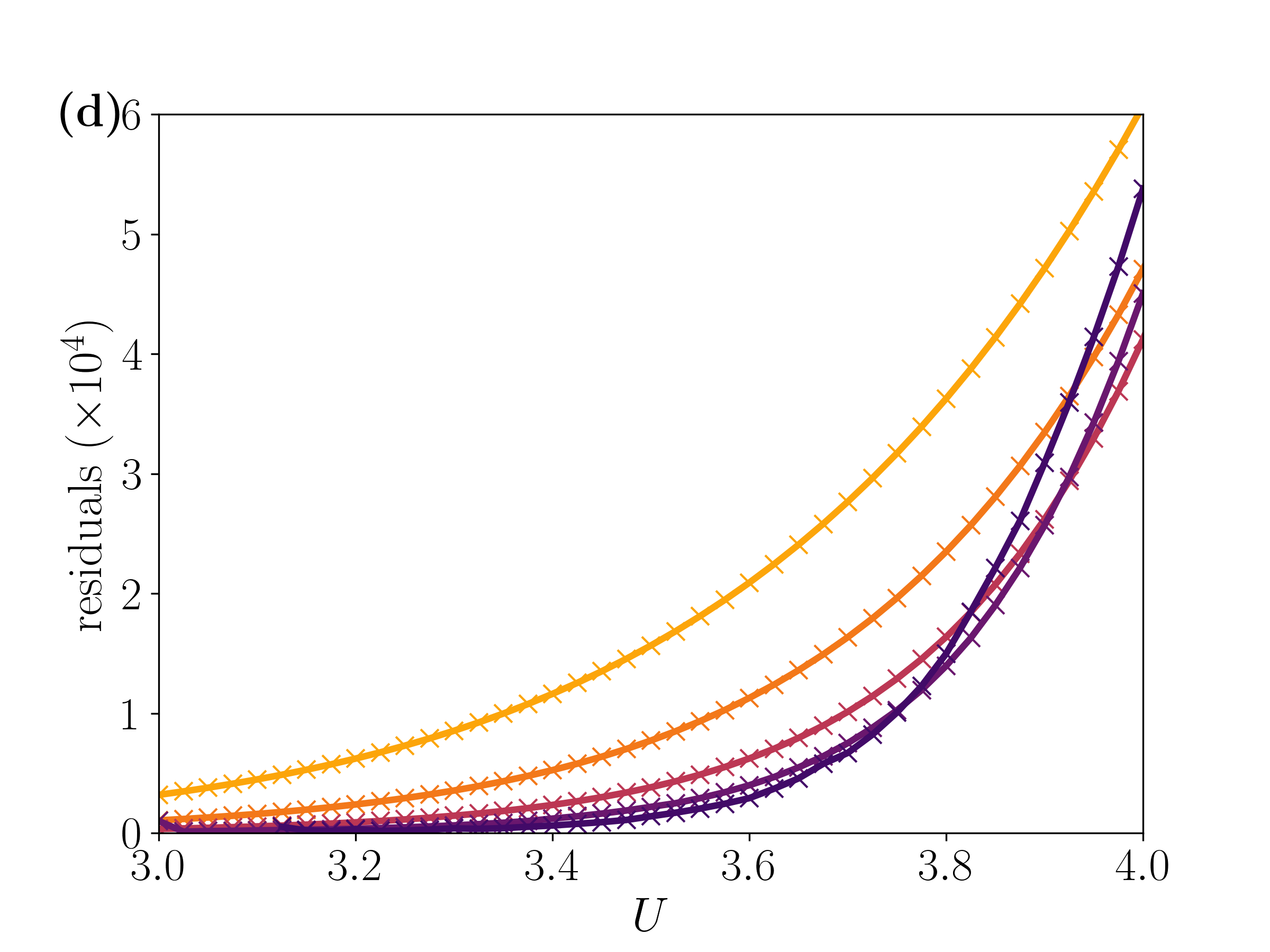}
    }
    \caption{\label{fig:cuts}(color online). Residuals of a gaussian fit of the PDF $p(m)\propto e^{- \beta m^2}$ for different system's sizes $L=16,32,64,128,256$ from light to dark color in the proximity of \textbf{(a)} the gapped-to-gapped phase transitions MI-HI-CDW along the cut $U=5$ and \textbf{(d)} the gapless-to-gapped phase transitions SF-MI (BKT) for $V=0$. In the insets, the location of the residuals' minimum is fitted against the inverse of the system's size in order to extrapolate the critical $V_C$ for the MI-HI phase transition \textbf{(b)} and the HI-CDW transition \textbf{(c)}.
    }
\end{figure*}
%%%%%%%%%%%%%%%%%%%%%%%%%%%%%%%%%%%%%%%%%%%%%%%%%%%%%%%

\section{Gaussian fit of the density PDF:}
For gapless one-dimensional phases described by a CFT, the PDF is a Gaussian, $p(m) \propto \exp(-\beta m^2)$,
due to the presence of a single bosonic generator in the theory
~\cite{LeHurLong,Laflorencie_2014,EranPRL2018}. 
More precisely, the ES has been shown to be organized in equally spaced parabolas $\xi_{\alpha\equiv (k,\gamma)}^{(m)} = \xi_0 + k\xi_1 + \beta m^2$, where the index $\alpha=(k,\gamma)$ denotes the order of the parabola and the degeneracy of the eigenvalue respectively, $\xi_0$ is the lowest eigenvalue of the ES, $\xi_1$ the difference between the lowest and the second eigenvalue in the $m=0$ particle sector~\cite{lauchli_gapless}.
As we will detail below, we expect instead the PDF of gapped phases to exhibit sensible deviations from this picture~\cite{lauchli_gapped}.

From the operational perspective, the differences between Gaussian and other PDF profiles can be captured via the average residuals of a two-parameters linear fit, namely $1/N_m\sum_m (\log p(m) + \beta m^2 + \beta')^2$ with $N_m$ the number of fitted points and $\beta'$ a normalization-related parameter.
In Fig.~\ref{fig:phase_diagram} we plot (the log of) such quantity over the span $m\in[-4,4]$ for a fit performed upon the distribution values for $m\in[-2,2]$.
%In Fig.~\ref{fig:phase_diagram} we plot the average residuals over the span $m\in[-4,4]$ of a two-parameters linear fit $- \log p(m) \simeq \beta m^2 + \gamma$ performed upon the distribution values for $m\in[-2,2]$.
This simple procedure turns out to be very effective to map out the entire phase diagram even without prior knowledge of the phases and the PDF shapes to be expected. 
We find perfect agreement with previous studies~\cite{fazio2012,Maciej2021}, without resorting to any study of the correlation functions or gap nor to the use of machine learning techniques and -- even more importantly -- by using an experimentally very accessible quantity. 
For the interested reader, we also provide in~\cite{SuppMat} the same Fig.~\ref{fig:phase_diagram} computed from data of simulated state-of-the-art experiments in which we reconstruct the PDF and we fit it from snapshots of the system for the particle counting. The results show that with a feasible number of snapshots, it is in principle possible to achieve a resolution comparable to the numerical study.

It is worth noticing that our method properly identifies even the recently postulated SF+SS phase, which appears as a very noisy region in Fig.~\ref{fig:phase_diagram}. 
This is a consequence of the alternation of uniform density regions and density-wave ordered ones~\cite{macejiSS}: The PDF is peaked on a value which is not necessarily the average density but depends on the location of the supersolid domains~\cite{SuppMat}. 
Since the PBC let this domains emerge in random positions along the ring, this gives rise to the above-mentioned noise. 

%%%%%%%%%%%%%%%%%%%%%%%%%%%%%%%%%%%%%%%%%%%%%%%%%%%%%%%
\begin{figure*}[t]
\centering %\hspace{-0.7cm}
    \begin{minipage}[t]{0.37\textwidth}
    \subfigure{
        \includegraphics[width = \textwidth]{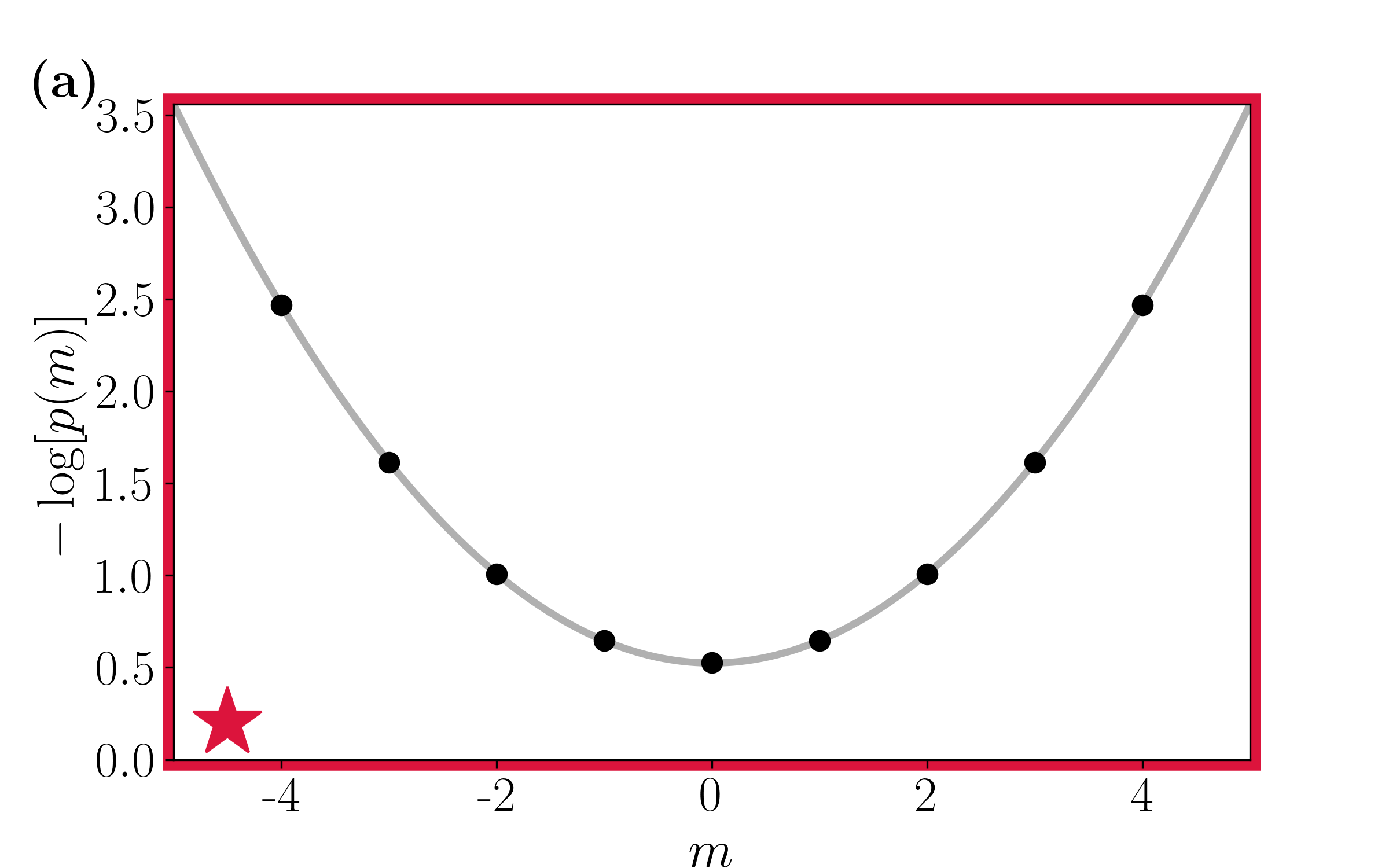}
    }\\ \vspace{-0.75cm}
    \subfigure{
        \includegraphics[width = \textwidth]{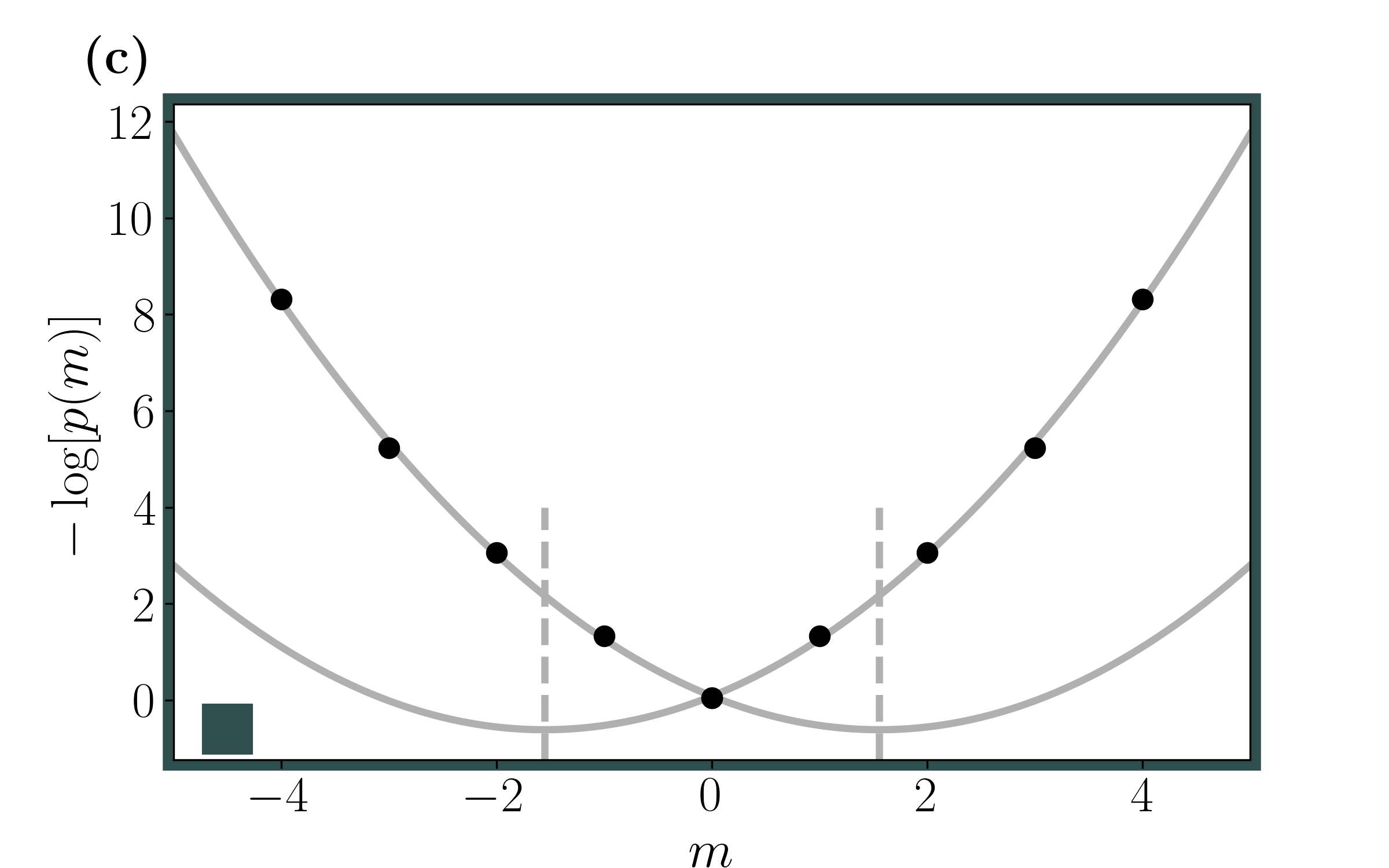}
    }
  \end{minipage}\hspace{-0.7cm}
  \begin{minipage}[t]{0.37\textwidth}
    \subfigure{
        \includegraphics[width = \textwidth]{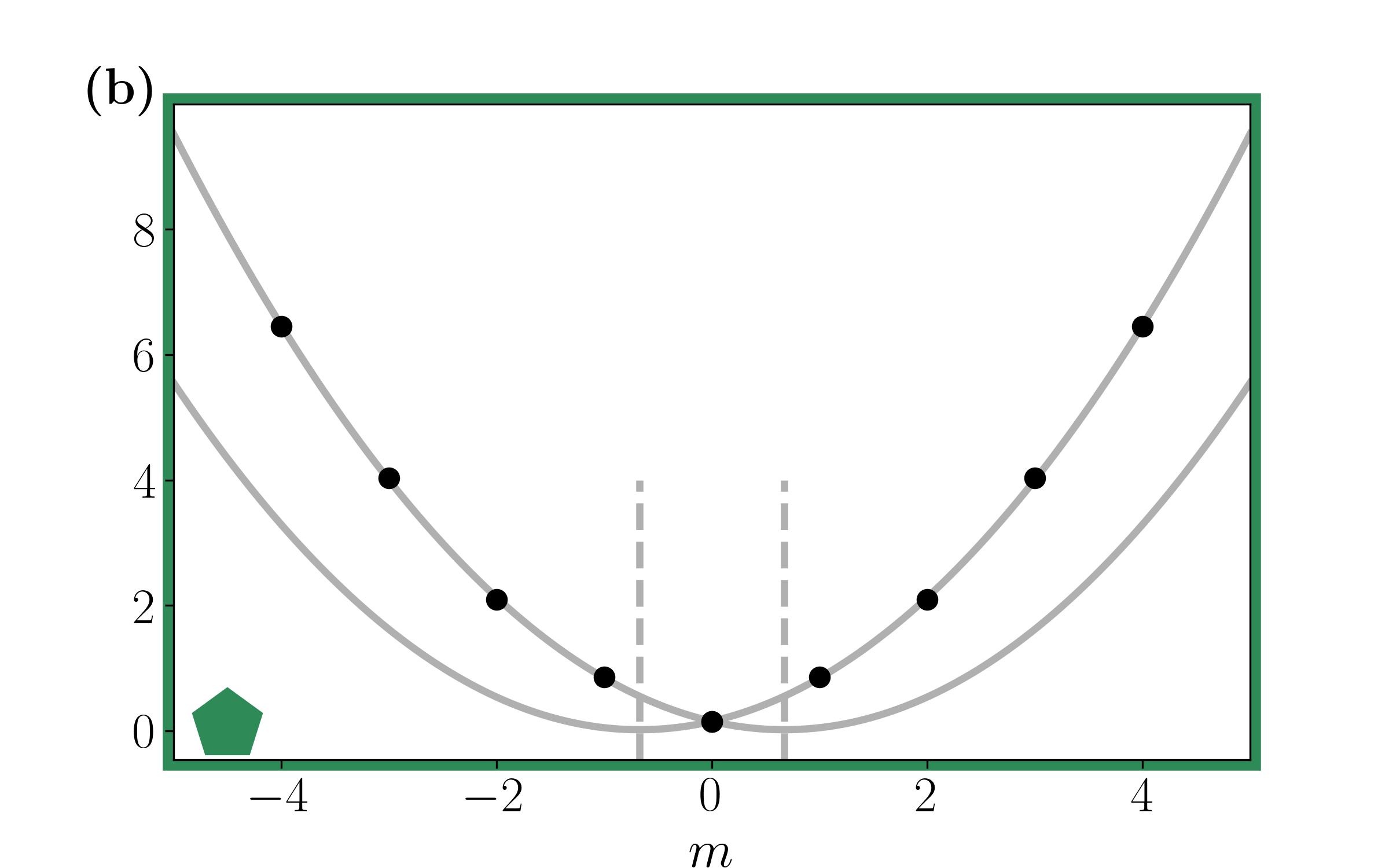}
    }\\ \vspace{-0.75cm}
    \subfigure{
        \includegraphics[width = \textwidth]{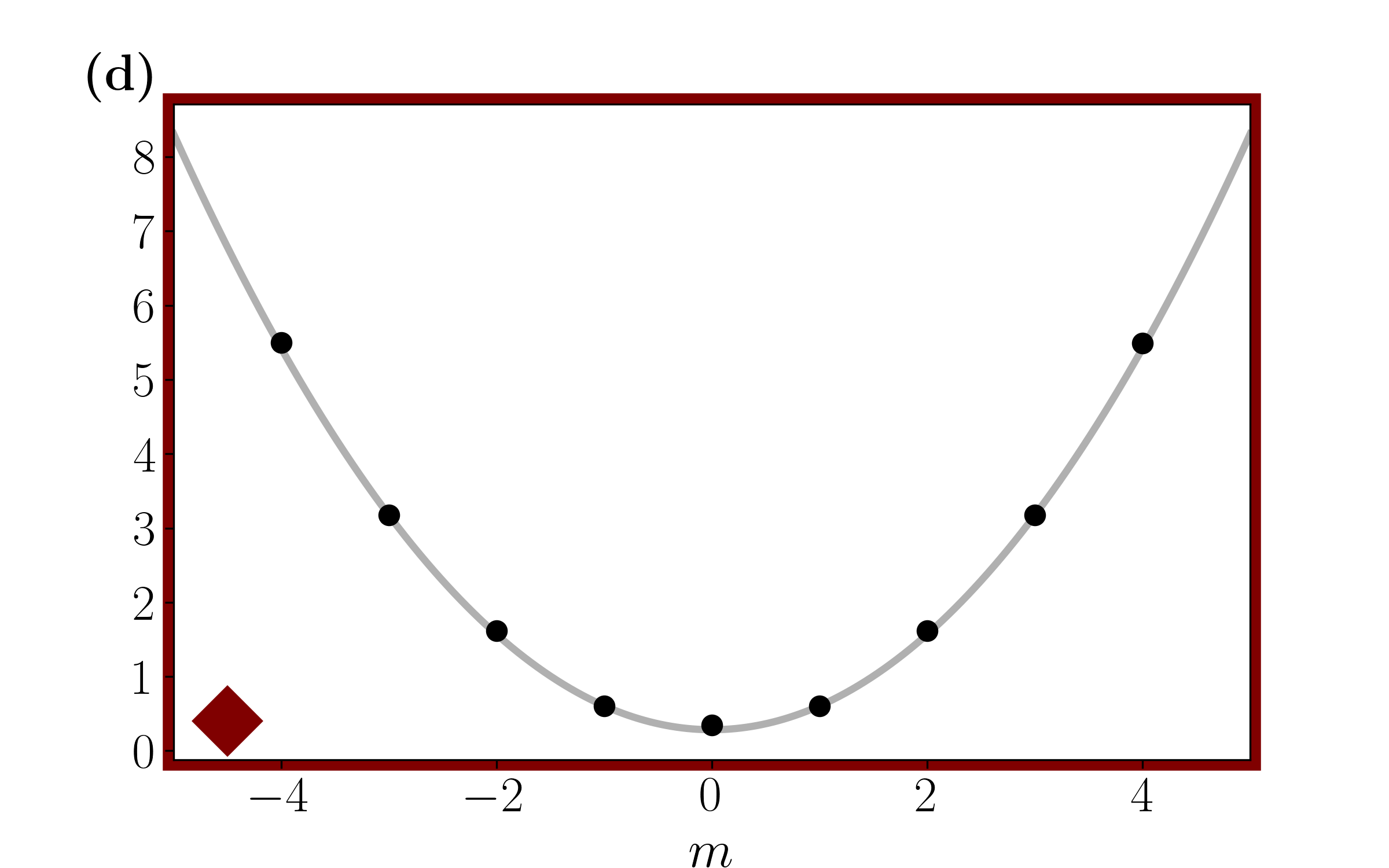}
    }
  \end{minipage}\hspace{-1cm}
  \begin{minipage}[t]{0.29\textwidth}\vspace{0.3cm}
    \subfigure{
        \includegraphics[width = 0.7\textwidth]{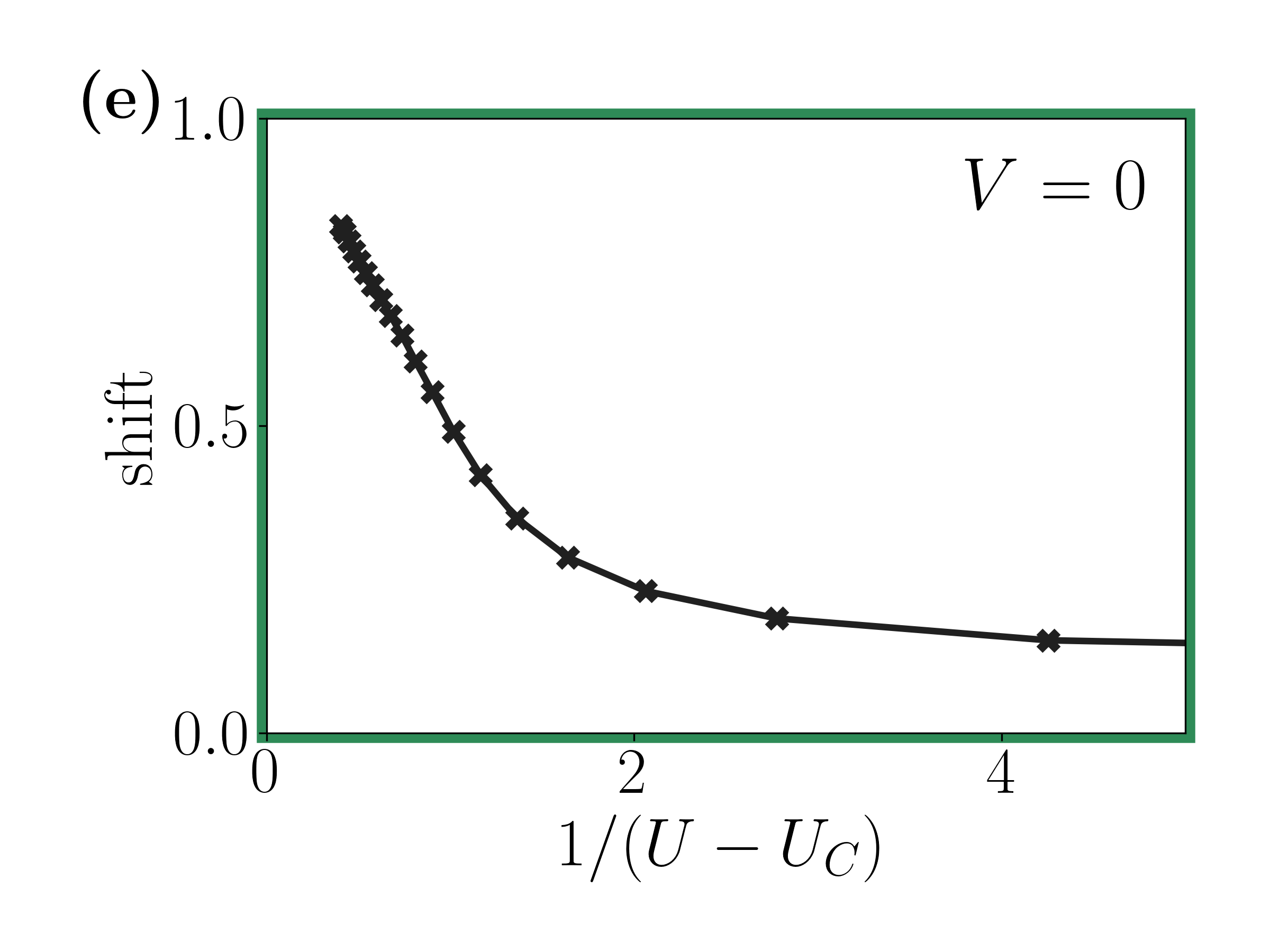}
    }\\ \vspace{-0.7cm}
    \subfigure{
        \includegraphics[width = 0.7\textwidth]{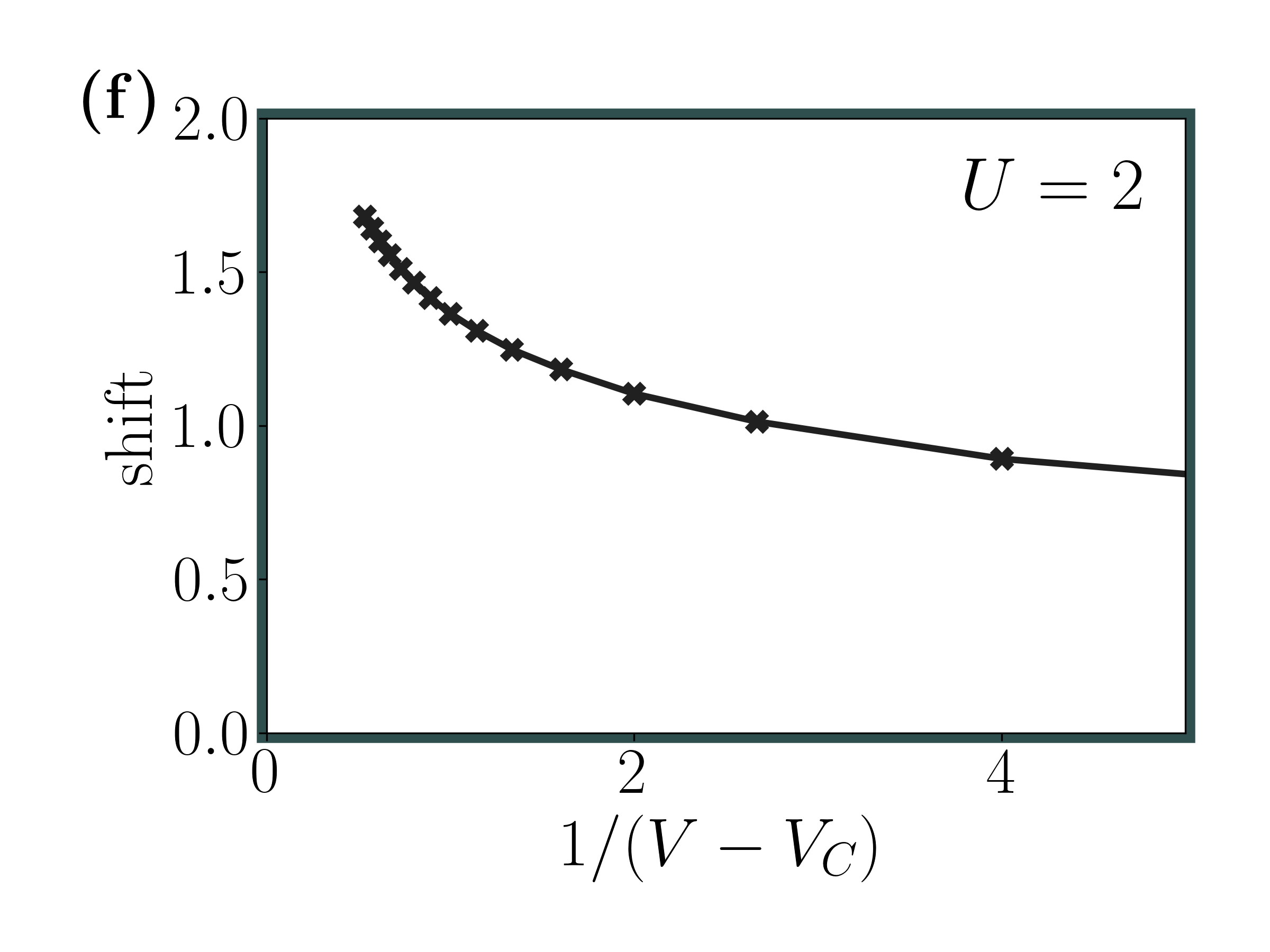}
    }\\ \vspace{-0.6cm}
    \subfigure{
        \includegraphics[width = 0.7\textwidth]{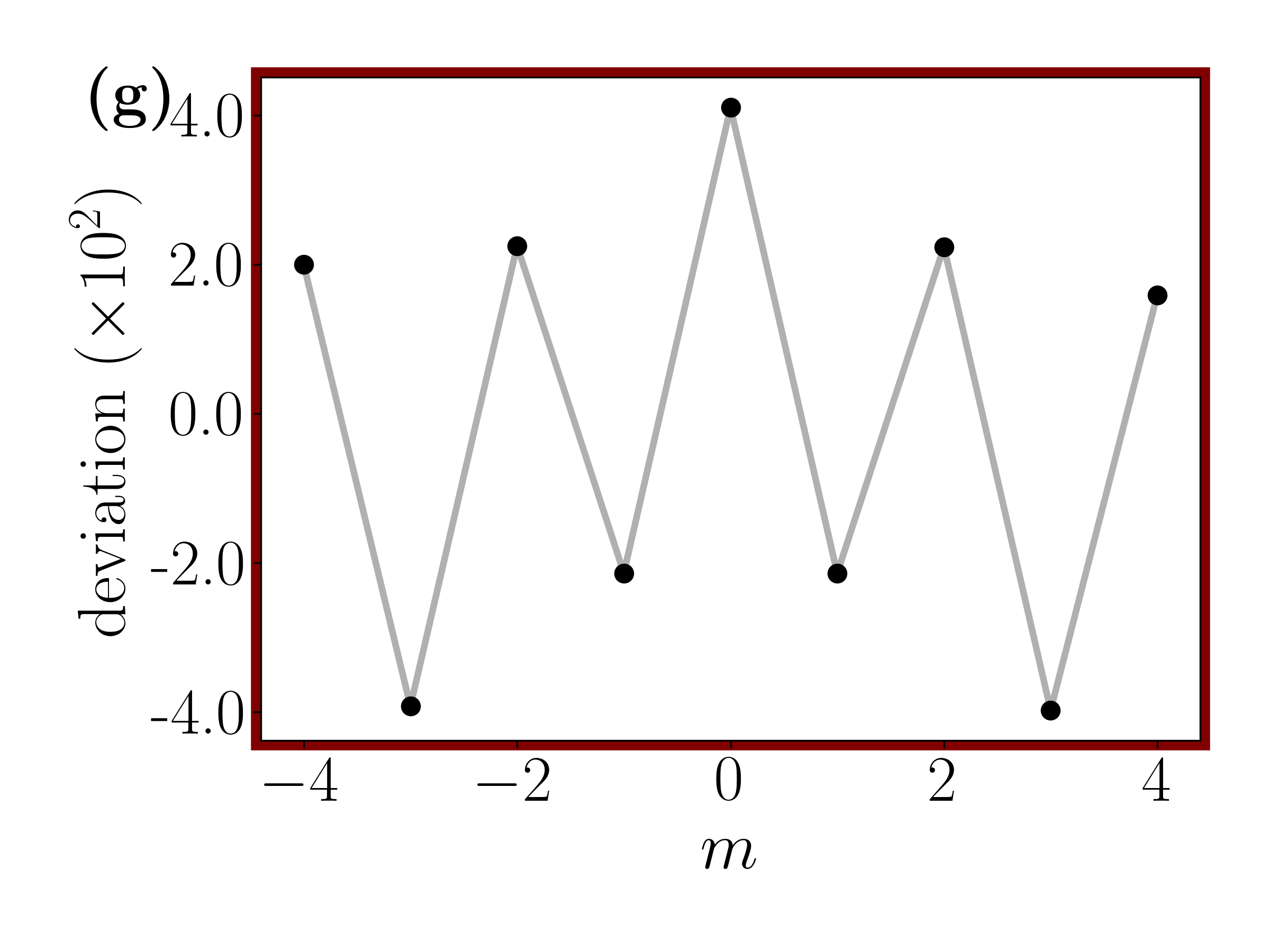}
    }
  \end{minipage}
    \caption{\label{fig:PDFdetails}(color online). The logarithm of the PDF (black points) is plotted for a representative configuration of the
        \textbf{(a)} SF phase (red star, $(U,V)=(0.5,0.5)$), 
        \textbf{(b)} MI phase (green pentagon, $(U,V)=(5.5,0.5)$), 
        \textbf{(c)} CDW phase (gray square, $(U,V)=(2,4.5)$), and 
        \textbf{(d)} HI phase (brown diamond, $(U,V)=(5,3.3)$). 
    The gray lines show the fit of the PDF in terms of Gaussian envelopes and their shifts, as discussed in the main text.
    For both MI and CDW, the behaviour is linear for small $|m|$, in compliance with horizontally shifted parabolas: on the right we show the shift of parabolas' minima from $m=0$ as a function of the inverse distance from the critical point for MI \textbf{(e)} and CDW \textbf{(f)}. It is apparent that they tend respectively to $1$ and $2$ deep in the phases, as predicted by perturbation theory in the text.
    For the HI, the dominant feature is a vertical shift between parabolas for the different parities of $m$: \textbf{(g)} shows a magnification thereof.
}
\end{figure*}
%%%%%%%%%%%%%%%%%%%%%%%%%%%%%%%%%%%%%%%%%%%%%%%%%%%%%%%

Moreover, the nature of the phase transitions is reflected in the evolution of the residuals in their vicinity: In Fig.~\ref{fig:cuts} we show the trend of the average residuals for two cuts of the phase diagram and different system's sizes ($L=16,32,64,128,256$ from light to dark color). 
%
%The nature of the phase transitions is reflected in the evolution of the residuals in their vicinity: the BKT between the MI and the SF has a quite smooth increase as well as from SF to HI. 
%On the contrary, the residuals allow an undeniable precise detection of the gapped-to-gapped transitions by identifying critical points as dips which divide regions of different residuals levels.
%
The first cut of Fig.~\ref{fig:cuts} \textbf{(a)} contains two gapped-to-gapped phase transitions for fixed $U=5$, from MI to HI and from HI to CDW as indicated with dotted line in Fig.~\ref{fig:phase_diagram}. 
%In contrast with the BKT transition, 
The location of the critical points is easily determined by the abrupt decrease of the residuals which becomes sharper and sharper as the length of the ring increases: 
In the two insets Fig.~\ref{fig:cuts} \textbf{(b-c)} a detail on the scaling of the critical $V_C$ against the inverse of the system size is shown together with a linear fit. 
The extrapolation of the critical values is in perfect agreement with the previous studies, i.e., $V=2.95\pm0.05$ for the transition MI-HI and $V = 3.525\pm0.05$ for HI-CDW~\cite{fazio2012}.

The second cut Fig.~\ref{fig:cuts} \textbf{(d)} encompasses the gapless-to-gapped BKT phase transition from the SF to the MI at $V=0$, indicated as vertical dashed line in Fig.~\ref{fig:phase_diagram}. 
It would be interesting to find a scaling procedure for the residuals, similarly to what is performed for the superfluid stiffness and/or the $K$-Luttinger parameter in standard approaches: At the moment, this remains however an open problem.
%\Teo{\emph{(eventualmente, mettere qui una frase ponte con lo scaling di dove lo shift delle parabole va a zero)}}
%
We stress, however, that the $\beta$ parameter of the linear fit offers a direct experimental access to the $K$-Luttinger parameter~\cite{LeHurLong}. For completeness, we perform such an analysis in~\cite{SuppMat} (see, also, Ref.\cite{Monien,GiamarchiBook} therein) and we obtain a critical value $U_c\simeq3.36\pm 0.01$ in perfect agreement with previous studies as reported in~\cite{matteoNJP}. The location of the transition corresponds to the uprising of the residuals.

%%%%%%%%%%%%%%%%%%%%%%%
%%% GENERAL PDF
%%%%%%%%%%%%%%%%%%%%%%%

\section{Shifted Gaussian PDF for gapped phases:}
A more detailed look at the PDF unveils actually more information than the simple residuals to gaussian fitting discussed above.
In Fig.~\ref{fig:PDFdetails} we present some prototypical configurations in the different phases, corresponding to the coloured symbols in the phase diagram of Fig.~\ref{fig:phase_diagram}: 
e.g., panel \textbf{(a)} shows the perfect parabola for the SF phase.
The striking feature is that all PDF profiles can be captured by proper shifts and combination of Gaussian envelopes, as we explain here below. 
%red star ($(U,V)=(0.5,0.5)$) in the SF phase

The shape of the PDF deep in the gapped phases can be computed resorting to boundary-linked perturbation theory~\cite{lauchli_gapped}. 
For example, the ES levels for the MI are obtained by consecutive applications of 
the kinetic term $H' = \sum_{j=1}^{L} - t (b^\dagger_{j+1}b_j + \mathrm{h.c.})$ on the zero-order ground state, i.e., $\vert\ldots 111 \ldots \rangle$.
We are then interested primarily in achieving a given unbalance $m$ with the minimal number of moves (i.e., perturbative orders of $H'$).
In the case of a single boundary, it is rather easy to see that the leading order amounts to $(t/U)^{ \frac{|m|(|m|+1)}{2}}$, up to a global weight depending on the ratio between the accumulated bosonic factors and the excitation energies along all possible sequences of moves~\cite{SuppMat}.
When dealing with two boundaries, simple combinatorics leads to $p(m) \simeq \log(t/U) \, \left\lfloor \frac{\left(|m|+1\right)^2}{4} \right\rfloor$.
The expression describes a symmetric envelope around the average number of particles, which could be recovered also by considering two symmetrical Gaussians shifted by $\mp1$ with respect to $m=0$, for the positive and negative values of $m$, respectively.
Remarkably, we find such horizontally shifted envelopes to persist with reduced offset when approaching the critical point, finally merging back when transitioning to the superfluid phase, see Fig.~\ref{fig:PDFdetails} \textbf{(b-e)}.
%green pentagon ($(U,V)=(5.5,0.5)$) in the MI regime
%The shift of the parabola's minimum as a function of the inverse distance from the critical point for $V=0$ (in inset of \textbf{(b)}) goes toward 1 moving deeper and deeper in the Mott phase as expected from the perturbative result. 
%
We also highlight here that a similar shift is the dominating feature of the PDF for the CDW phase, this time tending to $\pm2$ deep in the perturbative regime, hinting at the underlying structure of the zero-order ground state, i.e., $\vert\ldots 0202 \ldots \rangle$, see panel \ref{fig:PDFdetails}\textbf{(c-f)}.
%\Teo{\emph(sarebbe bellissimo riuscire a far tornare il conto, ma al momento non ci riusciamo piu'\ldots)}
% dark gray square ($(U,V)=(2,4.5)$)
%The representative of the CDW (dark gray square, \textbf{(c)} ($(U,V)=(2,4.5)$)) behaves in a similar way but the scaling in the inset shows that it approaches 2 far deep in the lobe. 

The appearance of the $|m|$ dependence in the exponent due to a non-zero shift of these double parabolic envelopes can be exploited to distinguish such gapped phases from all the others. 
We verified that the values of the $\alpha$ coefficient of the fit $-\log[p(m)] = \alpha |m| + \beta m^2 + \gamma$ is non-zero only in the MI and CDW. The change rate of the coefficient along the phase transitions manifests again their nature~\cite{SuppMat}. 
We also stress that for the single boundary partition of some specific models, instead, the PDF may assume asymmetric shapes, see a very nice description for the XXZ open chain in~\cite{lauchli_gapped}.

Finally, a direct inspection of the PDF in the HI phase reveals the typical footprint of the topological order. 
A detailed description can be obtained by truncating the maximum site occupation to $n_{max}=2$ bosons and mapping the EBH model to a spin-1 Heisenberg model~\cite{Spin1XXZSanctuary2003,HIDallaTorre2008,Holger2014HI}. 
In this framework the configurations of the HI appear as a \textit{dilute anti-ferromagnet} of doblons(2) and holons(0) separated by an undetermined number of single occupations(1), $\vert\ldots 2 1 \ldots 1 0 \ldots 2 1 \ldots 1 0 \ldots \rangle$~\cite{HIDallaTorre2008}. 
%
%brown diamond ($(U,V)=(5,3.3)$) in the HI phase
Noticeably, the PDF of this topologically gapped phase, panel~\ref{fig:PDFdetails} \textbf{(d)}, resembles very closely the Gaussian PDF of the gapless phase,
with the only signature of a slight shift for even/odd Gaussians (see Fig.~\ref{fig:PDFdetails} \textbf{(g)}), alluding to the underlying parity-string order parameter. 
%\Teo{\emph{(speriamo sia convincente abbastanza, anche se il conto non lo abbiamo)}}
%
%The logarithm of the PDF form a pure, centered parabola with the addiction of a small correction. In order to try grasping the quality of such a correction, we plot the deviation from the parabolic fit in the inset. 
%The topological order affects dissimilarly the probability for even or odd values of $m$.
%\Teo{\emph{(se sopra gia' spieghiamo il comportamento di HI, qui conviene dire che ci troviamo in accordo, anziche' riabbozzare di nuovo, no?)}}
%
This explains the sensibly reduced, though still sizeable, residuals in Fig.~\ref{fig:phase_diagram}. 

%%%%%%%%%%%%%%%%%%%%%%%
%%% CONCLUSIONS
%%%%%%%%%%%%%%%%%%%%%%%

\section{Conclusions:}
In conclusion, we have shown that the probability density function of the occupation number of a portion of the system (and simple fits thereof) can be a powerful agnostic inspection tool for the phase diagram of quantum many-body systems.
For the extended Bose-Hubbard model, the obtained results are comparable with much more sophisticated analysis both via traditional methods dealing with gap scalings and correlation functions~\cite{fazio2012} and via modern machine learning approaches fed with the ES~\cite{Maciej2021}.
We claim that the PDF provides the best intermediate quantity between the whole ES and the bipartite number fluctuations~\cite{LeHurEE} also for automatic detection protocols, since it combines the advantage of being easily accessible in modern experiments (e.g., quantum gas microscopes) with the wealth of information about the full many-body state, while requiring little prior knowledge about the emerging phases.
%for the purpose of detecting the phase diagram
Moreover, we foresee the method to be valid more in general: 
An extension beyond zero-temperature regimes and one-dimensional systems will be an appealing follow-up of this work.
%
%\Teo{\emph{(check ref LeHur! Esiste una review di gas microscopes in generale?)}}

%%%%%%%%%%%%%%%%%%%%%%%
%%% ACKNOWLEDGEMENTS
%%%%%%%%%%%%%%%%%%%%%%%
\section{Acknowledgements:} We acknowledge support from 
the Deutsche Forschungsgemeinschaft (DFG), project grant 277101999, within the CRC network TR 183 (subproject B01), 
%the European Union (PASQuanS, Grant No. 817482), 
the Alexander von Humboldt Foundation, 
the Provincia Autonoma di Trento, 
from Q@TN (the joint lab between University of Trento, FBK-Fondazione Bruno Kessler, INFN-National Institute for Nuclear Physics and CNR-National Research Council) 
and from the Italian MIUR under the PRIN2017 project CEnTraL. 
We gratefully acknowledge discussions with M. Kiefer-Emmanouilidis and A. Haller.
The authors gratefully acknowledge the Gauss Centre for Supercomputing e.V. (www.gauss-centre.eu) for funding this project by providing computing time through the John von Neumann Institute for Computing (NIC) on the GCS Supercomputer JUWELS (grant NeTeNeSyQuMa) and JURECA (institute project PGI-8) at Jülich Supercomputing Centre (JSC).
The MPS simulations were run with a code based on a flexible Abelian Symmetric Tensor Networks Library, developed in collaboration with the group of S. Montangero (Padua).

%The MPS simulations were run on the JURECA and JUWELS Clusters at the Forschungszentrum Jülich, with a code based on a flexible Abelian Symmetric Tensor Networks Library, developed in collaboration with the group of S. Montangero (Padua).
%\Teo{\emph{(for the final version check if there is a specific string for project-no. on clusters)}}
\nocite{*}

% \bibliography{bibliography}% Produces the bibliography via BibTeX.
\balance
\providecommand{\noopsort}[1]{}\providecommand{\singleletter}[1]{#1}%
%

% \balancecolsandclearpage
\balancecolsandclearpage
% \documentclass[%
%  reprint,
%  superscriptaddress,
%  amsmath,amssymb,
%  aps,
%  prb,
% ]{revtex4-2}

% \usepackage[utf8]{inputenc}
% \usepackage{xr-hyper}
% \usepackage{hyperref}
% \hypersetup{
%     colorlinks=true,
%     linkcolor=blue,
%     filecolor=magenta,      
%     urlcolor=cyan,
% }
% \usepackage{graphicx}
% \usepackage[caption=false]{subfig}
% \newcommand{\Dani}[1]{\textcolor{red}{{\textbf{D:} } #1}}

% \begin{document}
% \widetext

% \makeatletter
\renewcommand{\theequation}{S\arabic{equation}}
\renewcommand{\thesection}{S\arabic{section}}
% \renewcommand{\thefigure}{S\arabic{figure}}
% \renewcommand{\bibnumfmt}[1]{[S#1]}
% \renewcommand{\citenumfont}[1]{S#1}

% \title{Phase Diagram Detection via Gaussian Fitting of Number Probability Distribution}
% %Full-counting statistics: a simple phase diagram detector

% \author{Daniele Contessi}
% \affiliation{Dipartimento  di  Fisica,  Universit\`a  di  Trento \&  INO-CNR BEC Center,  38123  Povo,  Italy;}
% \affiliation{Forschungszentrum J\"ulich GmbH, Institute of Quantum Control,\\
% Peter Gr\"unberg Institut (PGI-8), 52425 J\"ulich, Germany; }
% \affiliation{Institute for Theoretical Physics, University of Cologne, D-50937 K\"oln, Germany}

% \author{Alessio Recati}
% \affiliation{Dipartimento  di  Fisica,  Universit\`a  di  Trento \&  INO-CNR BEC Center,  38123  Povo,  Italy;}

% \author{Matteo Rizzi}
% \affiliation{Forschungszentrum J\"ulich GmbH, Institute of Quantum Control,\\
% Peter Gr\"unberg Institut (PGI-8), 52425 J\"ulich, Germany; }
% \affiliation{Institute for Theoretical Physics, University of Cologne, D-50937 K\"oln, Germany}

% \date{February 2023}

% \maketitle
%%%%%%%%%%%%%%%%%%%%%%%
%%% SUPPLEMENTAL MATERIAL
%%%%%%%%%%%%%%%%%%%%%%%
\section*{SUPPLEMENTAL MATERIAL}
\setcounter{figure}{6}
\setcounter{page}{1}
\setcounter{section}{0}
\setcounter{equation}{0}
\setcounter{table}{0}

%%%%%%%%%%%%%%%%%%%%%%%%%%%%%%%%%%%%%%%%%%%%%%%%%%%%%
\section{Phase separation}
\label{app:ss_phase_separation}
\begin{figure}[b]    \centering
    \includegraphics[width = 0.48\textwidth]{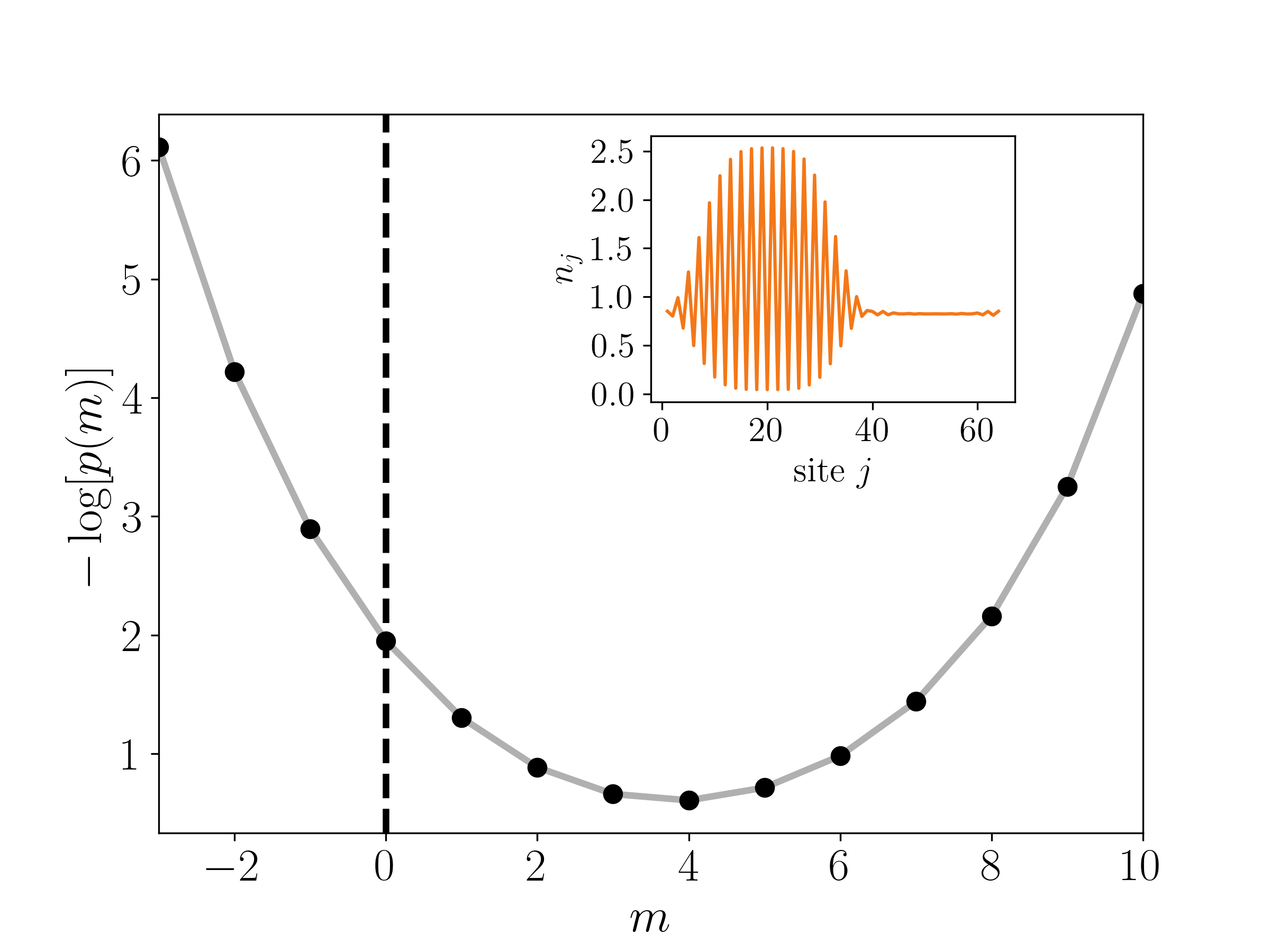}
    \caption{\label{fig::ss}(color online). In the main plot the PDF of a configuration in the phase separated regime SF + SS for a system of $L=64$. The inset shows the correspondent profile of the density.}
\end{figure}
In the bottom right corner of the phase diagram -- namely for weak on-site interaction $U$ and strong nearest neighbour one $V$ -- a phase-separated state between SF and supersolid (SF + SS) for unitary filling is present~\cite{Maciej2021_supp}. The residuals of a centered gaussian fit of the PDF appear there as very noisy. In order to show the origin of such a noise, we plot in Fig.~\ref{fig::ss} the PDF for $(U,V)=(0.5,4)$ of a system of length $L=64$ sites. The PDF is clearly not centered in $m=0$ in contrast with all the other phases we have discussed. Indeed, looking at the local density along the system in the inset, a supersolid domain has emerged with a periodic structure in the left region of the lattice while the right region is flat. A very detailed explanation about the phase separation can be found in~\cite{macejiSS_supp}.
%%%%%%%%%%%%%%%%%%%%%%%%%%%%%%%%%%%%%%%%%%%%%%%%%%%%%
\section{Perturbative calculation of states amplitudes in deep gapped phases}
As illustrated in the main text, in order to understand the origin of the modified envelopes for the $p(m)$ in the gapped phases, we carried out the calculation for of the leading contributions resorting to perturbation theory. Following the standard prescription with a perturbation $H' = \lambda W$ to the original hamiltonian, the relevant states at $k$-th order in our case are:
\begin{equation}
    \ket{\beta^{(k)}} \propto \lambda^k
    \frac{W_{\beta l_1}W_{l_1l_2}...W_{l_{k-3}l_{k-2}}W_{l_{k-2}\alpha}}{E_{\beta\alpha}E_{l_1\alpha}...E_{l_{k-2}\alpha}} \ket{\alpha^{(0)}} 
    %\frac{W_{nl_{k-1}}...W_{l_3l_2}W_{l_2l_1}}{E_{l_1n}E_{l_2n}...E_{l_{k-1}l_1}} \ket{l_1^{(0)}} 
\end{equation}
using the notation for the perturbation matrix elements and energies:
\begin{equation}
    \begin{split}
        W_{\beta\alpha}\equiv \langle \beta^{(0)}|W|\alpha^{(0)}\rangle ,\\
        E_{\beta\alpha}\equiv E_{\beta}^{(0)}-E_{\alpha}^{(0)}.
    \end{split}
\end{equation}
All the other terms are sub-leading because, for fixed $m$, require more than the minimum number of moves in order to transport the particles from from subsystem B to A (and vice-versa for negative $m$) and hence belong to a higher perturbative order. The recipe for the calculation of the full corrections, involves the summation of all possible combinations for moving $m$ particles though. Such a summation must be weighted with the ratio between the accumulated bosonic factors resulting from $\bra{\beta} \sum_{j=1}^{L} - t (b^\dagger_{j+1}b_j + \mathrm{h.c.})\ket{\alpha}$ and the energy.

For the single boundary case, by following the above mentioned protocol one finds that the amplitude of the states is proportional to $(t/U)^{ \frac{|m|}{2}\left( |m|+1 \right)}$ for the MI starting from the ground state $\ket{1,1,...,1,1}$ (in agreement with~\cite{lauchli_gapped_supp}).

The correspondent state for the two-boundary ES happens to have an amplitude proportional to 
\begin{equation}
    (t/U)^{\mathrm{Floor}\left(\frac{m+1}{2}\right)^2}
\end{equation}
% Instead, for the CDW the leading amplitude is proportional to
% \begin{equation}
%     \begin{cases}
%     (t/V)^{ \frac{|m|}{2}\left( \frac{|m|}{2}+2 \right)}\mathrm{ for }\ m\ \mathrm{ even,}\\
%     ...\\
%     \end{cases}
% \end{equation}
% Both for the MI and the CDW the envelopes for enven and odd values of $m$ are vertically close to each other and have the same horizontal shift. Indeed, by fitting the data with a unique parabola the values of the shift $\pm1$ for the MI and $\pm 2$ for the CDW are recovered deep in the respective gapped phase.  

\section{Linear dependence of the exponent of the PDF}
\label{app:gapped_pd}
\begin{figure}[b]    \centering
    \includegraphics[width = 0.48\textwidth]{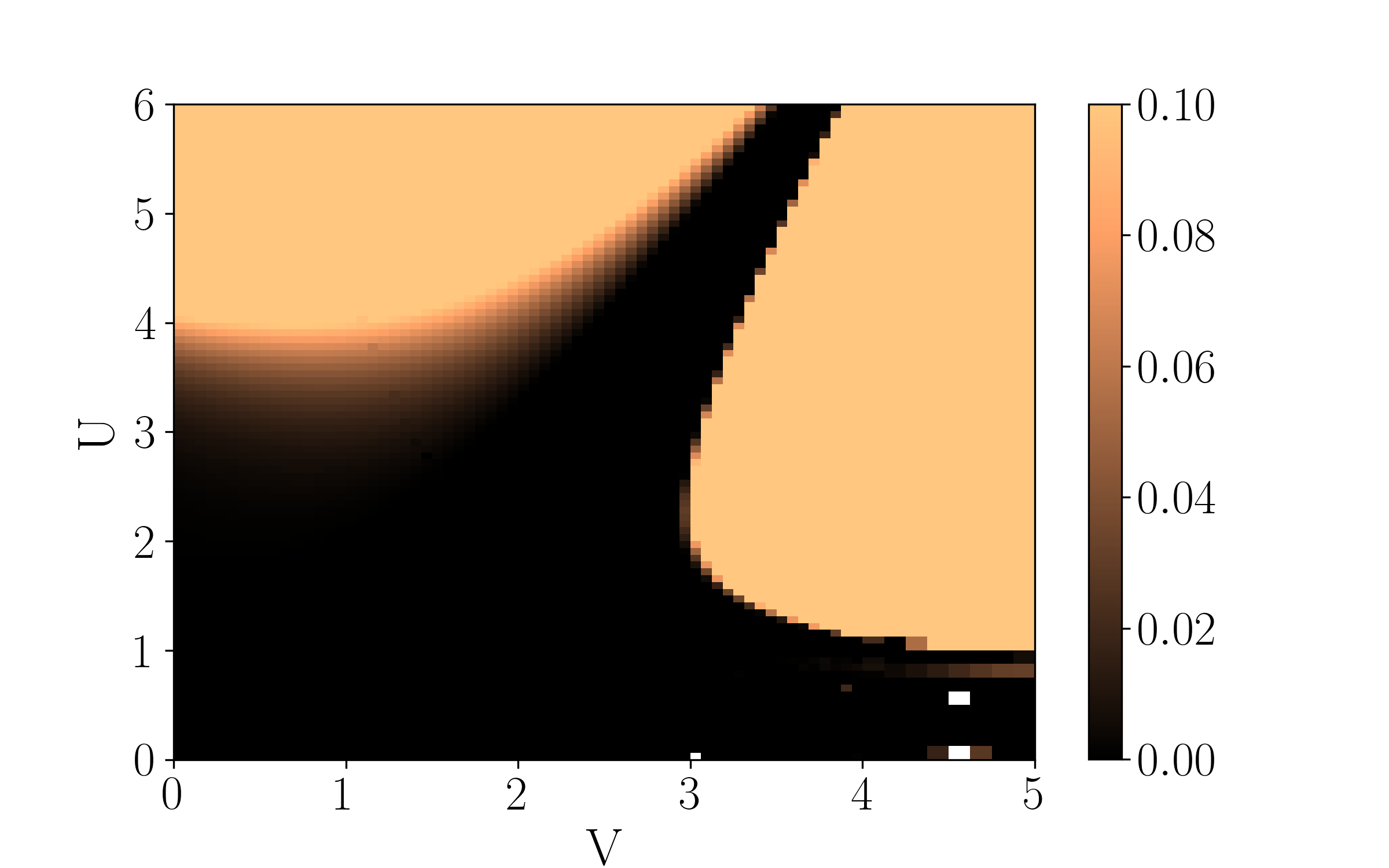}
    \caption{\label{fig::gapped_phase_diagram}(color online). Linear dependence of the logarithm of the PDF for small $m$ as a function of the interactions. }
\end{figure}
As mentioned in the main text, the linear dependence of the PDF for small $m$ can be used as a mark for the MI and CDW gapped phases. In Fig.~\ref{fig::gapped_phase_diagram} we plot the value of the $\alpha$ coefficient for the fit $-\log[p(m)] = \alpha |m| + \beta m^2 + \mathrm{const.}$ as a function of the couplings of the model. The non-zero value of this coefficient is the reason why a purely quadratic centered fit $p(m)\propto e^{-\beta m^2}$ fails in the above mentioned phases, giving rise to big residuals.
\begin{figure}[h!]    \centering
    \includegraphics[width = 0.48\textwidth]{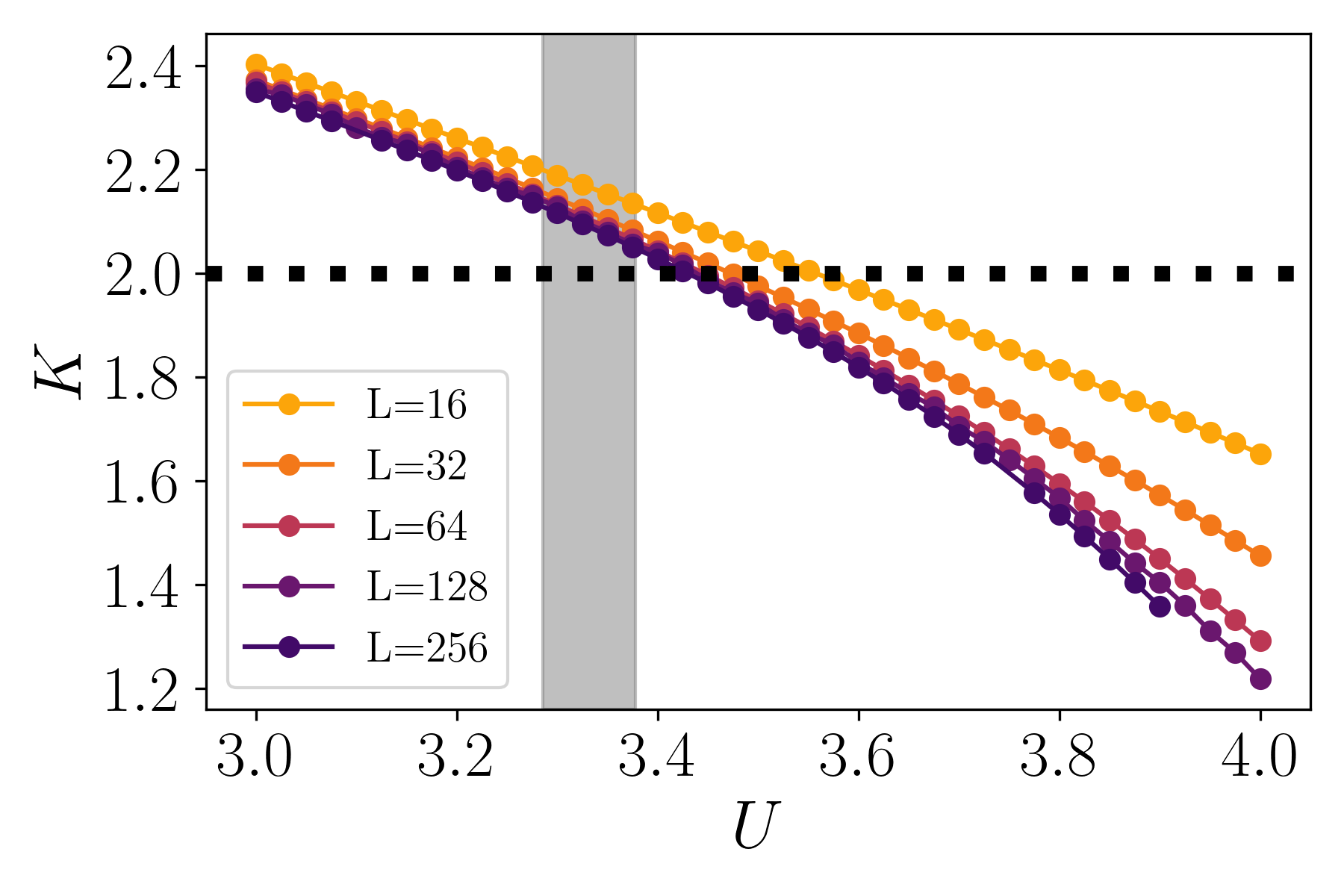}
    \caption{\label{fig::K_extrapolation}(color online). Detection of the BKT transition by means of the second moment of the PDF distribution, i.e. the number fluctuations (see~\cite{LeHurQCP_supp}). The curves of the effective Luttinger parameter $K^\ast$ values for different sizes of the system are shown as a function of $U$ ($t=1$ and $V=0$). The critical true value is expected at $K=2$ from the Luttinger theory (dotted black line). The gray stands for the single standard deviation confidence interval as from the estimates of literature summarized in Table 1 of~\cite{matteoNJP_supp}. }
\end{figure}
\begin{figure}[hb]    \centering
    \includegraphics[width = 0.48\textwidth]{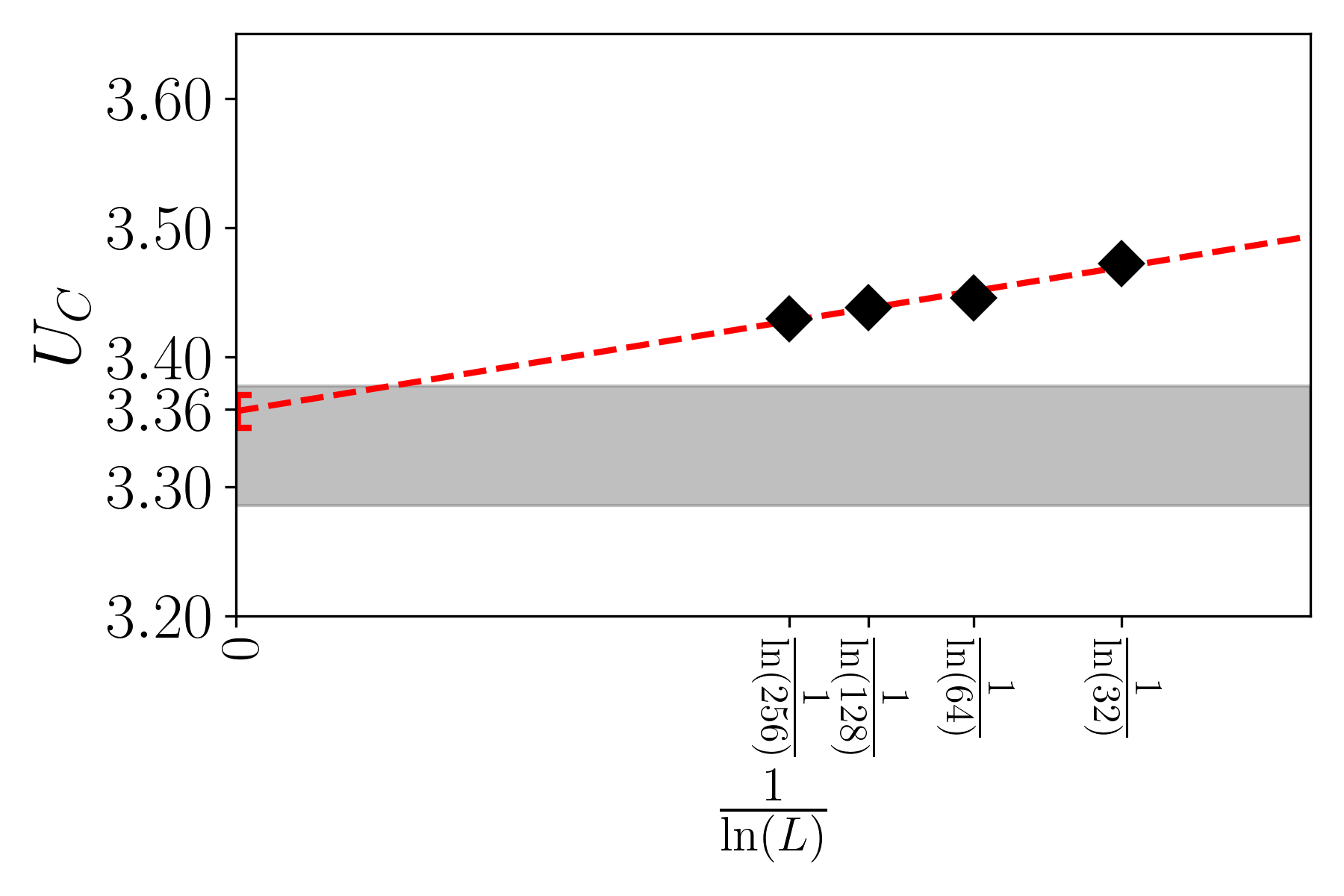}
    \caption{\label{fig::K_extrapolation_scaling}(color online). Extrapolation of the critical value: diamonds are the pseudo-critical values extracted from the crossing of the $K^\ast$ curves with $K=2$ of the previous figure, as a function of the reciprocal of the log-size. With a linear fit (red dashed line) the value of $U_C$ to the thermodynamic limit is estimated from the intercept.}
\end{figure}
\section{Precise detection of BKT transition from PDF}
In this section we perform the determination of the BKT phase transition at $V=0$ resorting to the expected critical value of the Luttinger parameter $K=2$ as predicted by the low-energy
hydrodynamic approximation~\cite{GiamarchiBook_supp}. This technique has already been used for pinpointing the value $U_C$ where the transition occurs via the extrapolation of $K$ either from the scaling of correlations~\cite{matteoNJP_supp,Monien_supp} or from the scaling of the charge fluctuations~\cite{LeHurLong_supp,LeHurQCP_supp,LeHurEE_supp}. Despite such a determination requires precise information about the physics of the transition and hence is not in the same agnostic spirit of our method presented in the main text, it is worth noticing that -- since the fluctuations correspond to the second moment of the PDF distribution -- it is also possible to compute $K$ directly from the PDF. The procedure simply takes into account the trend of the quadratic coefficient $\beta$ of the PDF versus the (log of) the bipartition length of the subsystem at fixed system size. In order to obtain the true value of $K$, one must perform an extrapolation to the thermodynamic limit therefore considering different system's sizes. We report the results of the aforementioned analysis in Figure~\ref{fig::K_extrapolation}: the effective $K^\ast$ per size of the system as function of $U$ is shown in the region of the BKT at $V=0$. As a guide-to-the-eye we plot with a gray shaded region the expected transition as from previous numerical studies listed in~\cite{matteoNJP_supp} and as black dotted line the expected critical value of the Luttinger parameter $K=2$. We derive the values of $U$ where the effective $K^\ast$ crosses with $K=2$ and we perform a fit against the inverse log of the system's size. This is represented in Fig.~\ref{fig::K_extrapolation_scaling}. Eventually we obtain the estimate of the critical coupling $U_C=3.36\pm0.01$ which is in agreement with previous studies.
\begin{figure*}[t]
    \centering
    \minipage{0.48\textwidth}
      \includegraphics[width=\linewidth]{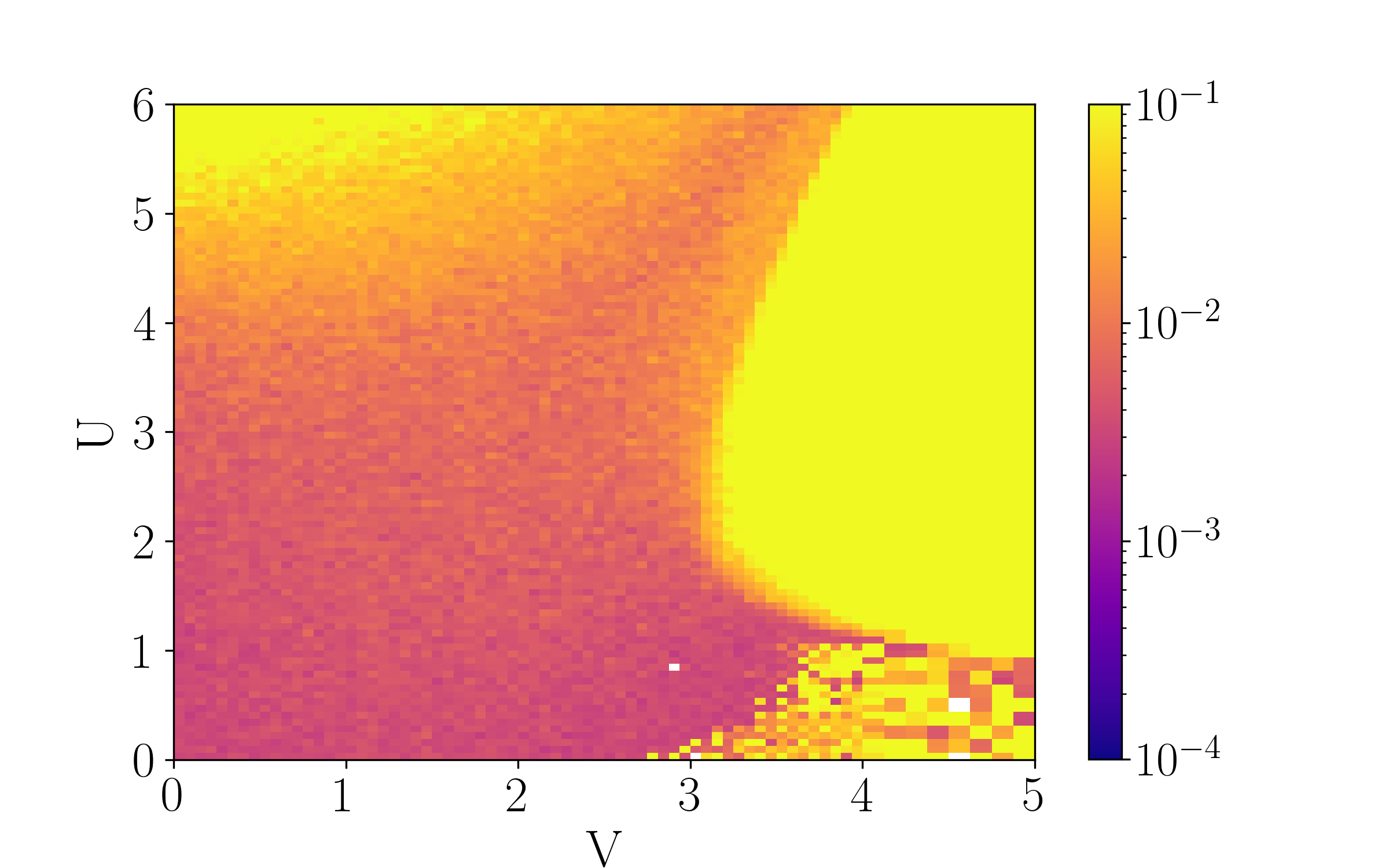}
    \endminipage\hspace{-0.9cm}
    \minipage{0.48\textwidth}
      \includegraphics[width=\linewidth]{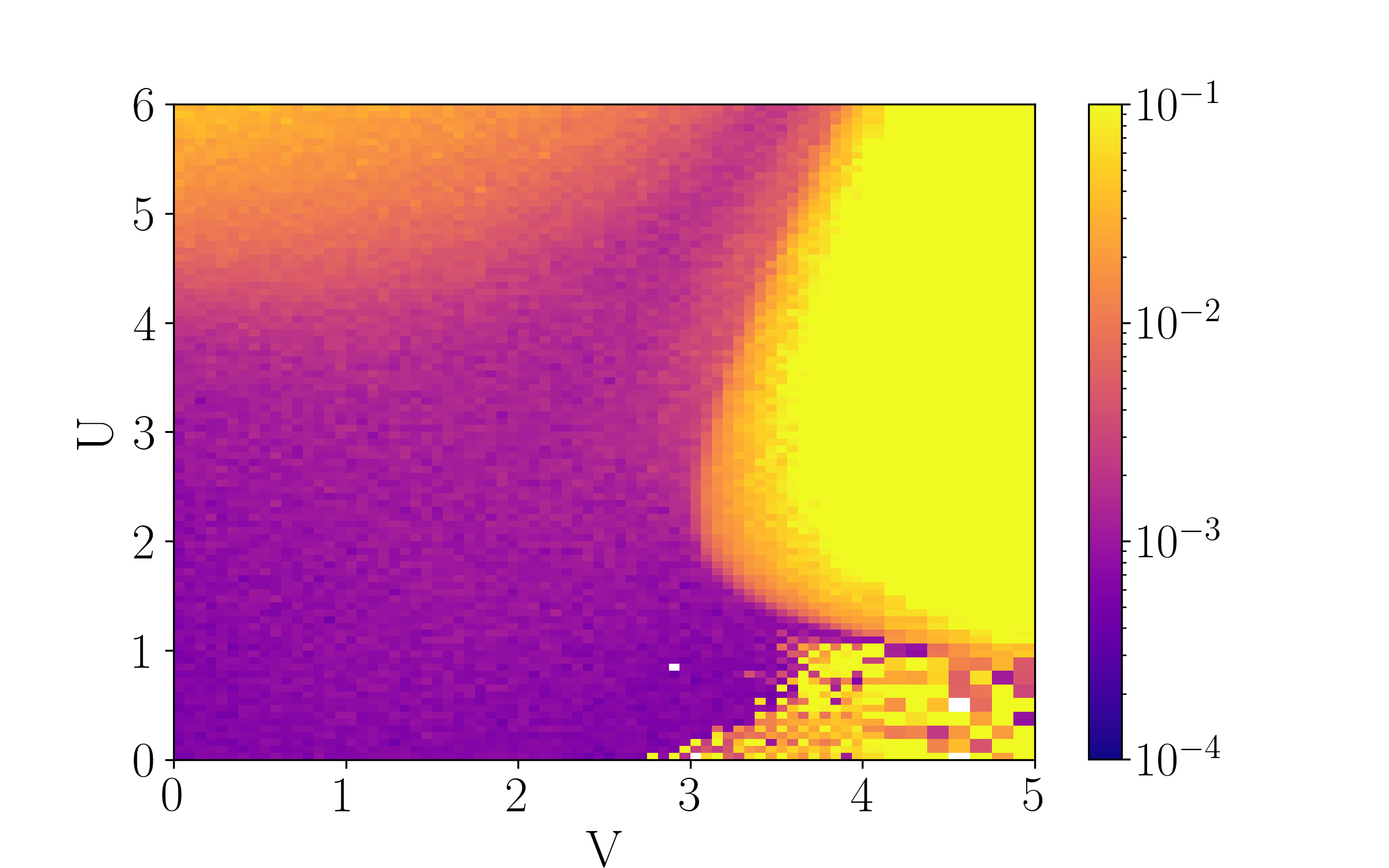}
    \endminipage\\
    \minipage{0.48\textwidth}
      \includegraphics[width=\linewidth]{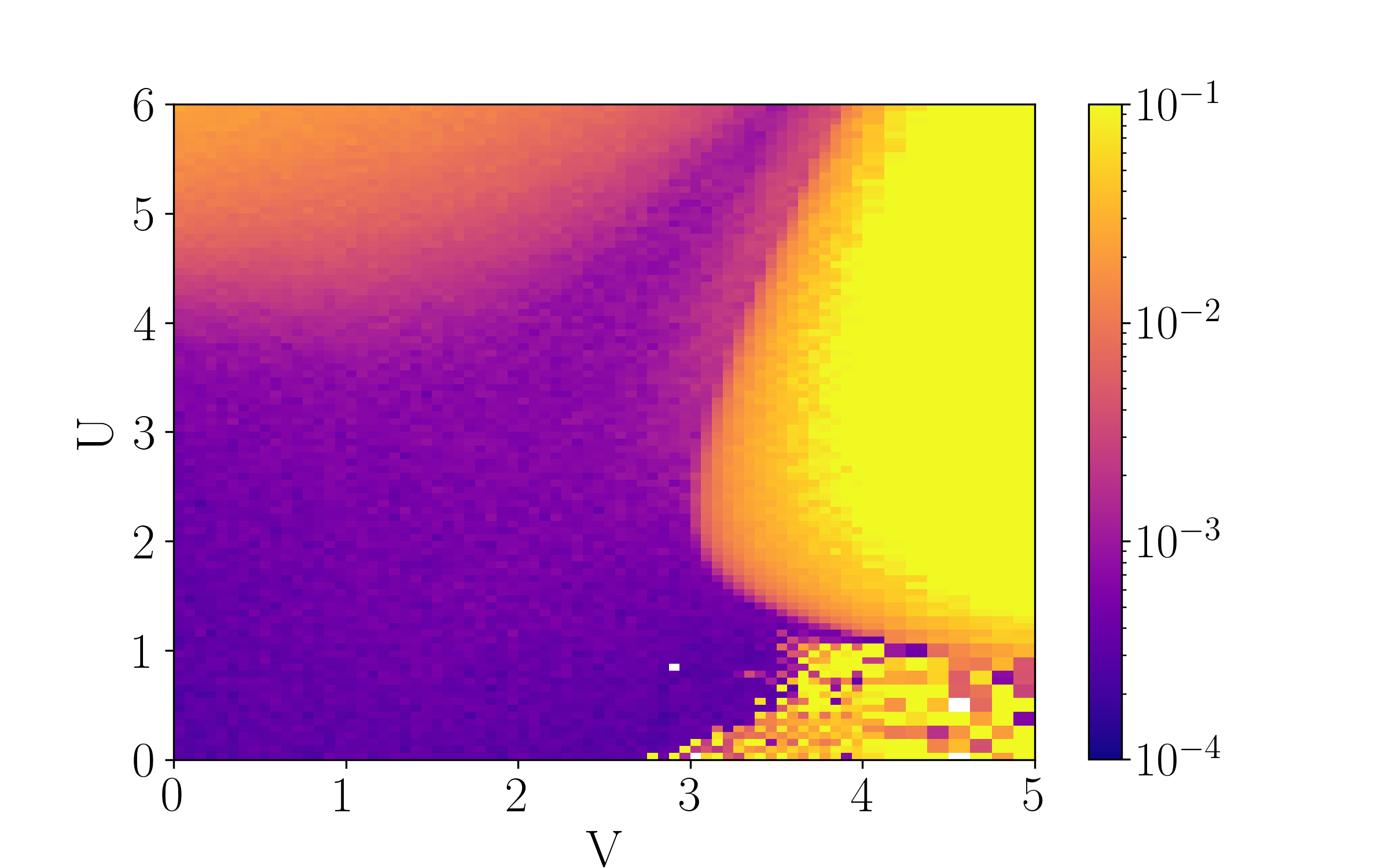}
    \endminipage\hspace{-0.9cm}
    \minipage{0.48\textwidth}
      \includegraphics[width=\linewidth]{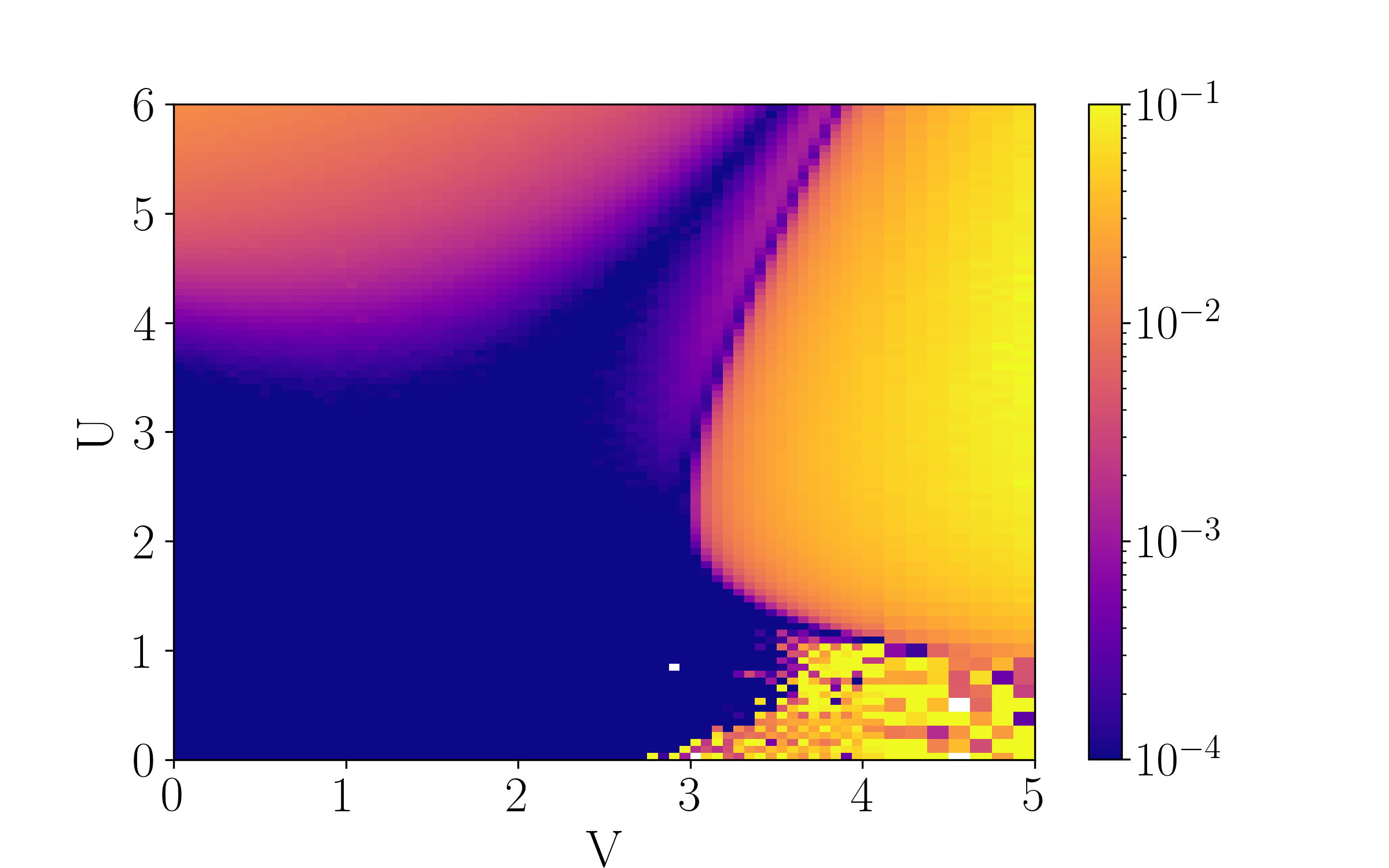}
    \endminipage
    \caption{
    \label{fig:simulated_exp} Phase diagram as of Fig. 1 of the manuscript obtained from different simulated experiments: the PDF is fitted from 100,500,1000,10000 shots (from left to right, top to bottom).}
\end{figure*}
\section{Phase diagram in simulated experiment}
In analogy with Fig.1 of the manuscript, we reproduce here the phase diagram of the EBH model by looking at the residuals of the gaussian fit on data of a simulated experiment. Instead of considering the numerical PDF of each pair of couplings $(U,V)$, we sample a given number of shots $N_S$ from it.
Every shot corresponds to a value for the unbalance in the number of particles $m$ among the two subsystems.
Once the samples are extracted, we fit the histogram of the sampled distribution and compute the residuals from there, therefore emulating an actual experiment that deals with the counting of the particles from the snapshots of the system.
The results are very promising, as shown in Fig.\ref{fig:simulated_exp}.
For a low number of samples ($N_S=100$ and $N_S=500$, top row) the only distinction between the deep MI and CDW phases is possible.
The phase separated SS-SF is already well distinguishable for the reasons explained in the main text and in the first section of the Supplementary Material.
Once one proceeds up to $N_S=10000$ shots (bottom right), the level of detail is good enough to even appreciate the HI and, in general, to be comparable to a numerical study.
Notice that those numbers of snapshots are normally achievable in state-of-the-art experiments with ultracold atoms on lattice since the current repetition time of an experiment is of the order of seconds. An hypothetical strategy could be to roughly draw the phase diagram with a few samples for every $(U,V)$ on a coarse grid and then refine the phases' borders with a greater (nonetheless feasible) number.

%%%%%%%%%%%%%%%%%%%%%%%
%%% BIBLIOGRAPHY
%%%%%%%%%%%%%%%%%%%%%%%
% \bibliography{bibliography}
\providecommand{\noopsort}[1]{}\providecommand{\singleletter}[1]{#1}%
%

% \end{document}
\end{document}